\documentclass[a4paper]{article}
 \pdfoutput=1

\usepackage{amssymb}
\usepackage{amsthm}
\usepackage{amsmath}
\usepackage{fullpage}
\usepackage{graphicx}
\usepackage{authblk}

\usepackage{lineno}

\begin{document}

\title{A study on boundary separation in an idealized ocean model}

\author{Peter D. D\"uben\thanks{dueben@atm.ox.ac.uk}}
\author{Peter Korn\thanks{peter.korn@mpimet.mpg.de}}
\affil{Max Planck Institute for Meteorology,\\Bundesstrasse 55, 20146 Hamburg, Germany}

\maketitle




\begin{abstract}

In numerical ocean models coast lines change the direction from one grid cell to its neighbor and the value for viscosity is set to be as small as possible.  Therefore, model simulations are not converged with resolution and boundary separation points differ in essential properties from flow separation in continuous flow fields. 
In this paper, we investigate the quality of the representation of boundary separation points in global ocean models. To this end, we apply well established criteria for boundary separation within an idealized ocean model setup. We investigate an eddy-resolving as well as a steady test case with idealized and unstructured coast lines in a shallow water model that is based on a finite element discretization method. 

The results show that well established criteria for separation fail to detect boundary separation points due to an insufficient representation of ocean flows along free-slip boundaries. Along no-slip boundaries, most separation criteria provide adequate results. However, a very sophisticated criterion based on dynamical system theory reveals that the representation of boundary separation is limited for these flows as well. 
We conclude that the representation of boundary separation points in numerical ocean models is not satisfying. This will have an impact on the separation of boundary currents in global ocean models. 

\end{abstract}

Keywords: boundary separation; boundary currents; no-slip boundary condition; free-slip boundary condition; global ocean models; dynamical system theory


\section{Introduction}
\label{sec:int}

Boundary separation belongs without doubt to the classical subjects of Fluid Dynamics. Since Prandtl's pioneering work in 1904 many studies were performed over the decades. As a result, Prandtl's theory for flow separation has been extended and criteria for flow separation are available for unsteady flows from no-slip and free-slip boundaries, that have proven to provide good results both in continuous flows and in numerical experiments. 
On the other hand, boundary separation of ocean currents, such as the Gulf Stream or the Kuroshio, were studied in physical oceanography for many decades but a satisfying theory that explains the position and variability of separation points is still not available, mainly due to the multitude of processes that influence separation in the ocean, such as changes of the wind stress, potential vorticity, pressure gradients, interactions with deeper currents, surface cooling and topography (see \cite{Chassignet2008} for an overview).

{\it The perspective of this paper is to test if boundary separation along the coast line is represented realistically within global circulation models of the ocean.} In global ocean models, the coastline is unstructured with changing alignment from one grid cell to its neighbor and the model solution is not converged with resolution since viscosity is set to be as small as possible. In contrast to many applications in Computational Fluid Dynamics, flow separation points along the boundaries are hardly resolved in these models. It is an interesting question if separation criteria that have been derived for continuous fluids will be applicable in simulations of ocean models.
If separation criteria are not applicable, this provides insight on how bad the representation of boundary currents that follow the coast line actually is. 

A multitude of papers discuss how separation of the Gulf stream is influenced if certain parameters in the model setup of global, three-dimensional ocean models are changed and what should be changed in the model setup to improve the representation of boundary separation. This paper does not continue this discussion since no changes will be applied to model simulations to influence separation. 
Furthermore, the results of this paper do not provide numerical evidence for the validity of the investigated separation criteria that have already been tested rather extensively in the literature of computational fluid dynamics. Previous studies can and will not be invalidated by our results.

We test the applicability of criteria for boundary separation in two idealized ocean setups. We reduce the ocean model to the shallow-water equations with no topography and study separation in two dimensions to avoid three-dimensional interactions that might influence the results and make the evaluation more complicated. The first setup that investigates boundary separation from idealized, straight coast lines is not eddy-resolving. The second setups is eddy-resolving with unstructured coast lines.   
We use a finite element shallow-water model for which we can assume that it will have a better representation of the coastline and the investigated boundary conditions (no-slip and free-slip) compared to many ocean models that are used today since it avoids staircase pattern along the coastline that are present in finite difference models and since it provides a clear representation of the physical fields at each point in space. In finite difference schemes the effective boundary conditions can actually dependent on the angle between the coast line and the coordinate axis of the numerical grid (see \cite{Adcroft1997} for the analysis on an Arakawa C-grid and B-grid).

We will use two well-known separation criteria to identify boundary separation points that are discussed in ocean modeling, namely Prandtl's separation condition (\cite{Prandtl1904}) and the adverse pressure gradient (\cite{Haidvogel1992}). We will than extend the discussion with two comparably new approaches that aim to treat boundary separation with innovative methods from the mathematical fields of topology and dynamical system theory that are introduced in the following paragraphs. 

Ghil, Ma, and Wang studied the topology of flow fields in a series of papers and presented a theory to determine structural bifurcations of two-dimensional incompressible vector fields (\cite{Ma2001, Ghil2001, Ghil2004, Ghil2005}). They investigated mainly no-slip, but also free-slip boundary conditions. The study provides criteria for the emergence of a new separation 
point, together with a reattachment point, from a flow field parallel to the boundary and should be able to detect separation points when they develop along the coast line. The criteria are dependent on vorticity.

Haller and colleagues (see \cite{Haller2004}, \cite{Lekien2008}
and references therein) studied boundary separation from the perspective of dynamical system theory for no-slip as well as for free-slip boundary conditions. For the no-slip case in an idealized 
picture, material lines  that are aligned with the boundary form a spike at the position of a separation point. Particles from the vicinity of the boundary are transported to the interior of the domain along this 
``material spike formation'', this is the definition of a separation point used in this approach. The flow trajectory at the separation point, 
is modeled as a non-hyperbolic saddle point. While Eulerian criteria, such as vanishing 
wall shear, can identify material spike formations in steady flows, it is known from numerical simulations and also 
from experiments that Eulerian quantities do generally not coincide with these points in unsteady flow fields 
(see for example \cite{Weldon2008}). For unsteady flows, material spike formations can not be identified by looking 
at instantaneous streamlines\footnote{A streamline is tangent to a snapshot of the velocity field at a given time 
and is different from a trajectory a particle takes in an unsteady flow field.}. 
Finally, necessary and sufficient criteria for separation points are derived
such that Prandtl's theory is recovered for steady flows. But since the theory of Haller and colleagues is also valid
for unsteady flows it can be seen as an extension of Prandtl's work to unsteady flows with fixed separation of 
two-dimensional flows from no-slip boundaries. In the case of free-slip boundary conditions the flow 
trajectories of separation points are assumed to be hyperbolic, and not non-hyperbolic as in the no-slip case. 
Lekien and Haller elaborated necessary and sufficient criteria for unsteady and moving separation points (\cite{Lekien2008}).

In section two, we give a very short description of the model setup for our simulations. In section three, we introduce the 
test cases. In section four, we present the criteria for flow separation that are applied to the test cases in section five. In section six, we draw conclusions.

\section{Model setup}

This section offers a brief introduction to the functionality of the model, including the shallow-water equations, the discretization in space and time and the post processing of the model data. A more detailed description of the model setup can be found in \cite{Dueben2012}.

\subsection{Viscous shallow-water equations and model discretization}

The used finite element model simulates the viscous shallow-water equations in non-conservative form

\begin{equation}
	\partial_t \mathbf{u} + \mathbf{u} \cdot \left( \nabla \mathbf{u} \right) + f \mathbf{k} \times \mathbf{u} + g \nabla h - \frac{1}{H} \nabla \cdot  \left( H \nu \nabla \mathbf{u} \right) = \frac{\boldsymbol{\tau}^s}{H } - \gamma_f \mathbf{u}, \notag
\end{equation}
\begin{equation}
\label{bo_shallowh}
 	\partial_t h + \nabla \cdot \left( H \mathbf{u} \right) = 0, \notag
\end{equation}
where $\mathbf{u}$ is the two-dimensional velocity vector, $f$ is the Coriolis parameter, $\mathbf{k}$ is the vertical unit vector, $g$ is the gravitational acceleration, $\nu$ is the eddy viscosity, $\boldsymbol{\tau}^s$ is the surface wind forcing, $\gamma_f$ is the bottom friction coefficient, $h$ is the surface elevation and $H$ is the height of the fluid column given by $H=h-h_b$, where $h_b$ is the bathymetry. 
The prognostic variables are surface elevation and velocity.

The used model can run with either free-slip ($\mathbf{u} \cdot \mathbf{n} = 0 $, and $\partial_\mathbf{n} \mathbf{u} = 0 $ on the boundary $\partial \Omega$), or no-slip boundary conditions ($\mathbf{u} = 0 $ on $\partial \Omega$). We introduce free-slip boundaries in a strong form and adjust the numerical fluxes through the boundaries. 
To introduce no-slip boundary conditions, we set the normal velocity flux at the boundary to zero and use a weak formulation through a penalty term to push the tangential component of velocity towards zero.

We use a $P_1^{DG}P_2$ finite element model. This means the model is using a discontinuous linear representation for velocity and a continuous second order representation for height. The time integration is performed with an explicit three level Adams-Bashforth method.

\subsection{Post processing of the model output}
\label{bo_postprocessing}

To identify separation points we evaluate physical fields of model simulations along the coast line. To investigate separation criteria based on dynamical systems theory, the physical quantities need to be defined continuously in space and time and not as discrete values, as in model output, to allow the calculation of flow trajectories. To this end, we use a bicubic spline interpolation in space, and a third order Lagrange interpolation in time to obtain a smooth representation of the diagnosed quantities, as recommended in \cite{Mancho2006}. 
Time interpolation is necessary, since we do not write model output after each time step, due to limitations in storage capacity. When a second spatial derivative in the tangential direction is needed, we calculate it by central differentiation along the coast line.

We use the original finite element representation of the velocity field -- which is changing linearly within one grid cells and is discontinuous between two adjacent grid cells -- to calculate particle trajectories along the boundary for free-slip boundary conditions. The discontinuous velocity field is smoothed in time by first order Lagrange interpolation. 
The use of higher order interpolation methods does not lead to a benefit, since the time difference between two model outputs is small enough. Corner points of the coastline between adjacent grid cells are not treated in a special way.

\subsection{Nomenclature for separation points}
\label{Sec:Turns}

\begin{figure}[ht!]
 \centerline{\includegraphics[width=0.45 \textwidth, angle=0]{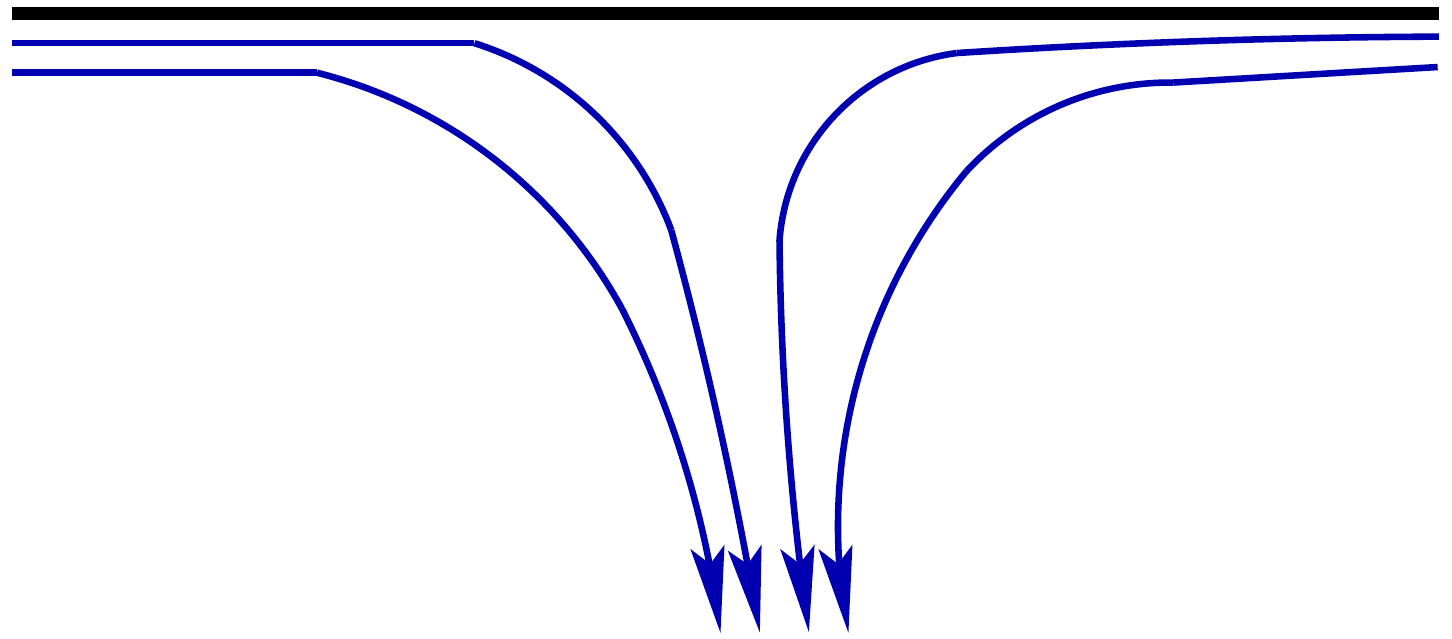} \includegraphics[width=0.45 \textwidth, angle=0]{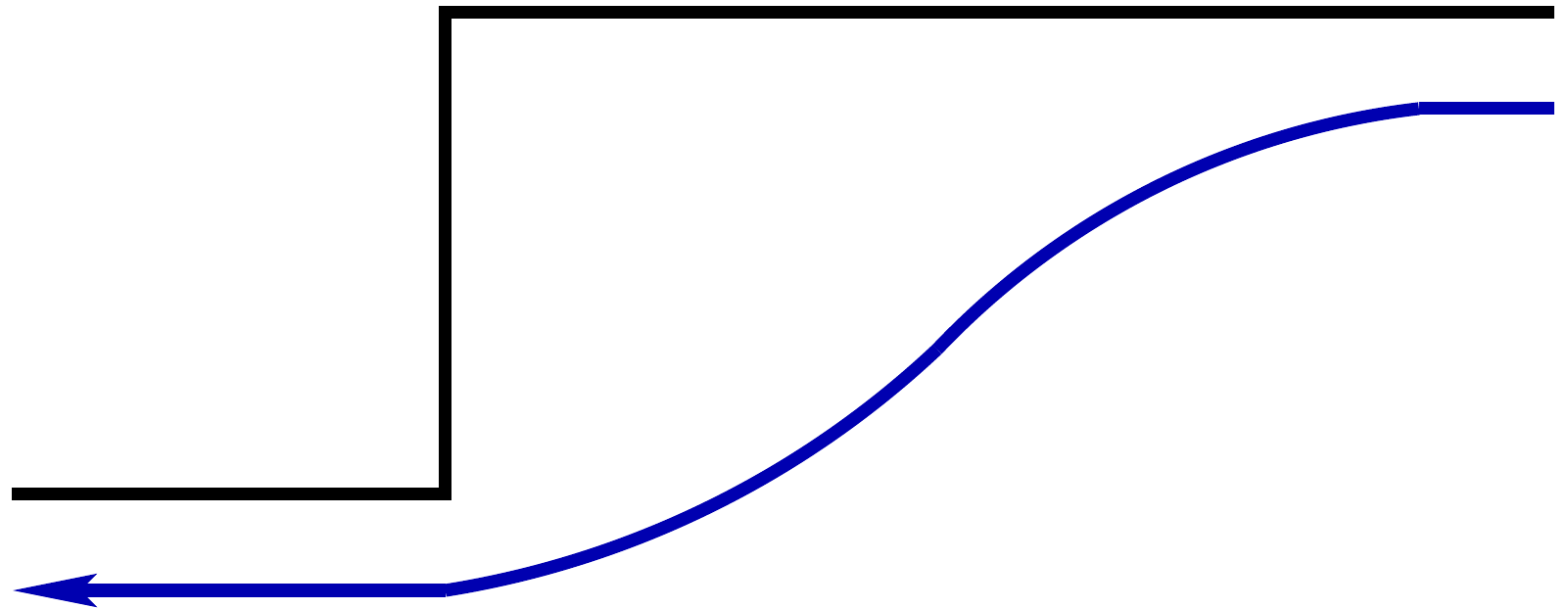} }
 \caption{Sketch of a clear separation point (left) and a flow around a turn of the coast line (right).}
  \label{bo_fixpoint_types}
  \end{figure}

 In this paper, we will distinguish between separation points and flows around turns of the coast line. The difference is sketched in Figure \ref{bo_fixpoint_types}. Points in which the flow separates from the coast line for more than one or two grid cells will be counted as a separation point and not as a turn of the coast line. It is arguable if flows that separate for less then two grid cells should be counted as separation points as well, but we will not.

\section{Test cases}
\label{bo_testcases}

In this section, we introduce the two test cases that are evaluated. In the idealized coast line test, 
we simulate a wind driven western boundary current that separates at the corner of an obstacle along straight coast lines. 
In the island test, we study the separation of a geostrophic flow around an island that has an unstructured coast line.
In contrast to the idealized coast line test, the island test is eddy-resolving.

\subsection{The idealized coast line test}
\label{bo_Dengg_TC}

We study an ocean gyre in the northern hemisphere. The gyre is forced by wind and rotates in clockwise 
direction. Due to the change of the Coriolis parameter in the meridional direction, the gyre is intensified 
towards the western boundary and a western boundary current develops (\cite{Stommel1948,Pedlosky1996}). 
The current separates from the coast at the edge of a rectangular obstacle. The setup is chosen to be as 
close as possible to the setup used in \cite{Dengg1993}, who investigated boundary separation in a barotropic vorticity model. 

\begin{figure}[ht!]
   \centering
   \includegraphics[width=0.6 \textwidth, angle=0]{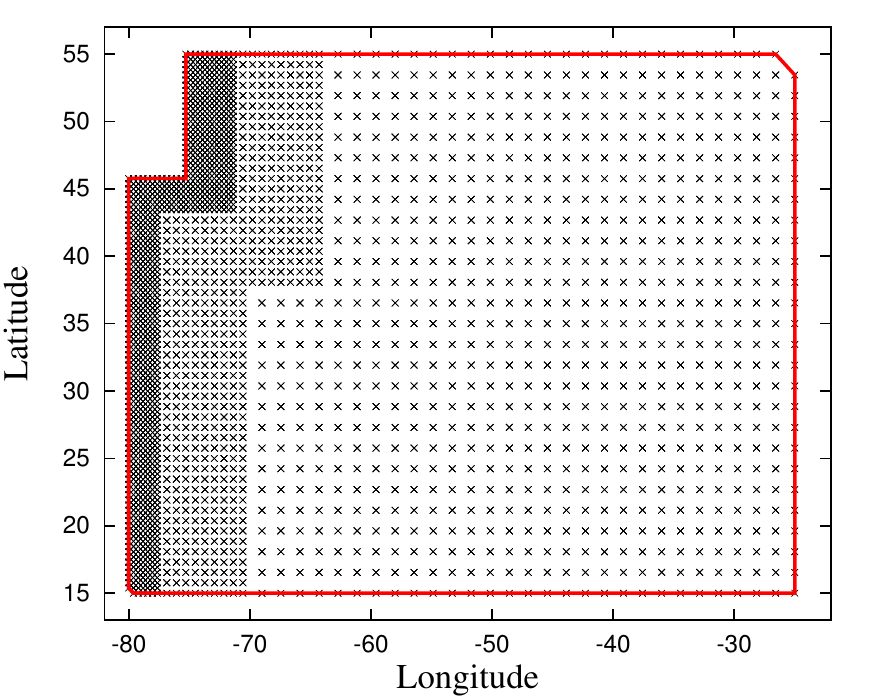}
 \caption{Vertices of the grid for the idealized coast line test. The continuous line marks the coast line. Grid refinement is used to increase the local resolution along the western boundary.}
  \label{bo_fig:dengg_grid}
  \end{figure}

We perform model runs on a planar, triangular grid which is structured in longitude/latitude space. Static h-refinement -- new grid points are introduced to the grid in regions of specific interest -- is used to increase the resolution at the boundary current (see Figure \ref{bo_fig:dengg_grid}). We performed tests with uniform high resolution to verify that the grid refinement does not influence the behavior of the boundary current. A grid edge has a length of about $1.6^\circ$ in the coarsest and $0.4^\circ$ in the finest part of the grid. A detailed investigation of grid refinement in the used finite element model can be found in \cite{Dueben2014}.

While the meridional wind forcing is zero, the zonal wind forcing is set to

\begin{equation}
 \tau^s_{\lambda} = \tau_0 \cdot 10^{-3} \cdot \cos \left( \frac{\pi \left(\theta - 15^\circ \right)}{40^\circ}  \right) , \notag
\end{equation}

where $\theta$ is the latitude. The bottom friction coefficient $\gamma_f$ is set to $10^{-6} \; s^{-1}$. The height field is initialized with a constant water depth of $1000 \; m$; the initial velocity is zero.

\begin{figure}[ht!]
   \center
\includegraphics[width=0.45 \textwidth, angle=90]{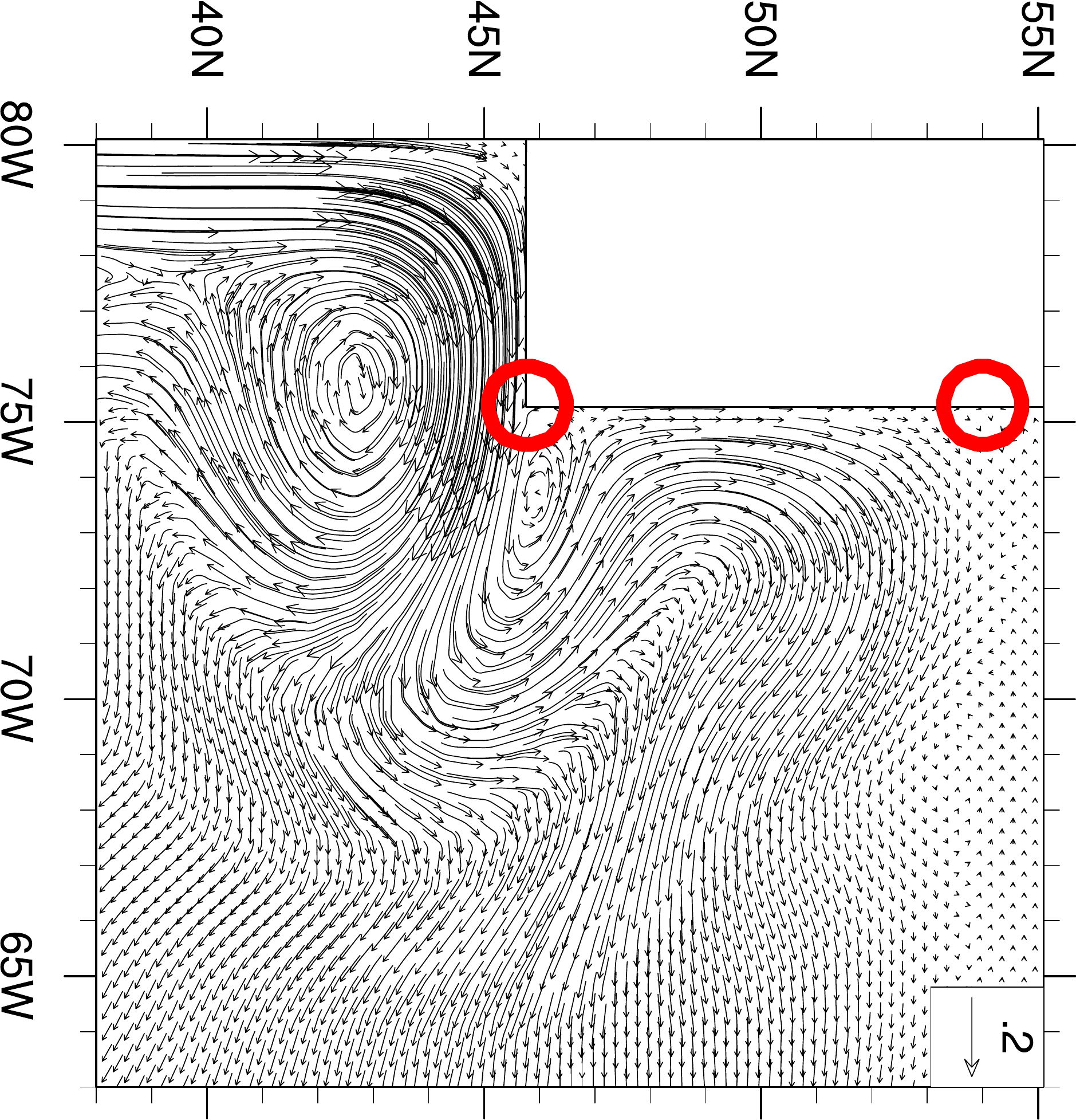} \includegraphics[width=0.45 \textwidth, angle=90]{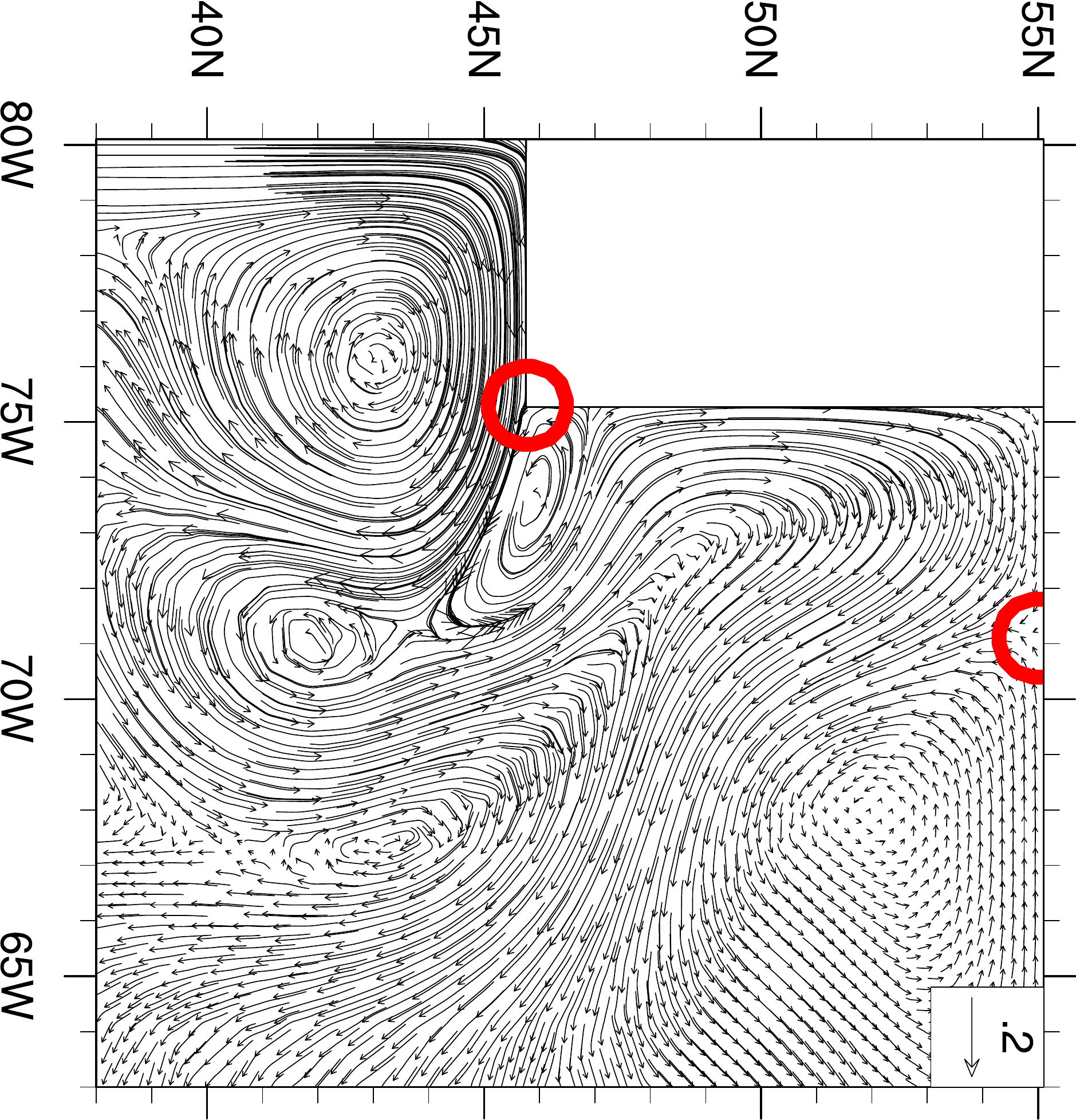} 
 \caption{Equilibrated velocity field in the upper left part of the idealized coast line test with no-slip (left) or free-slip (right) boundary conditions. The red circles mark separation points that were identified by looking at the flow field. 
In the no-slip run, the northern separation point might appear to be set too far to the north, but the flow trajectories in the direct vicinity of the coast line are separating at this point.}
  \label{bo_fig:dengg_fs_jk}
\end{figure}

Figure \ref{bo_fig:dengg_fs_jk} shows the equilibrated steady velocity field for $\tau_0 = 0.84 \; m^2 s^{-2}$, $\nu = 3000 \; m^2 s^{-1}$, and either no-slip, or free-slip boundary conditions in the upper left part of the domain, after one year of integration.

\subsection{The island test}
\label{bo_Dong_TC}

We study a geostrophic flow around an island. We simulate a global, zonal jet similar to the one in the steady-state zonal geostrophic flow test proposed in \cite{Williamson1992} (test case 2 with $\alpha =0$), and introduce a small island into the northern hemisphere. The simulations are performed on a global icosahedral geodesic grid. The physical fields are initialized as follows

\begin{align}
 u = u_0 \cos \left( \theta \right), \quad v = 0, \quad \text{and} \quad h = h_0 - \left( a_e \Omega u_0 + \frac{u_0^2}{2} \right) \frac{\sin \left( \theta \right)^2}{g}, \notag
\end{align}

where $\theta$ is the latitude, $a_e$ is the radius of the earth, $\Omega$ is the earth rotation rate, $u$ is the zonal velocity, $v$ is the meridional velocity, and $g$ is the gravitational acceleration. We choose $u_0 = 2.83 \; ms^{-1}$ and $h_0 = 500 \; m $.

The island has a diameter $D$ of 110 km and is centered around $45^{\circ}$ North and $0^{\circ}$ West. Since we simulate the whole globe, although we are only interested in the flow around the small island, we refine our numerical grid extensively. We start from a coarse icosahedral grid with an averaged edge length of 960 km, and introduce seven refinement levels, where each level reduces the lattice spacing by a factor of two. 
A typical edge length at the boundary of the island is 7.5 km. The island is cut out of the grid by removing all grid points within a specific distance to the center of the island.

The Reynolds number of the flow around the island is given by

\begin{equation}
  Re = \frac{UD}{\nu}, \notag
\end{equation}

where $D$ is the diameter of the island, $\nu$ is the viscosity, and $U$ is a typical value for velocity, which is 2 $ms^{-1}$ in the vicinity of the island. 

\begin{figure}[ht!]
   \center
  \includegraphics[width=0.4 \textwidth, angle=90]{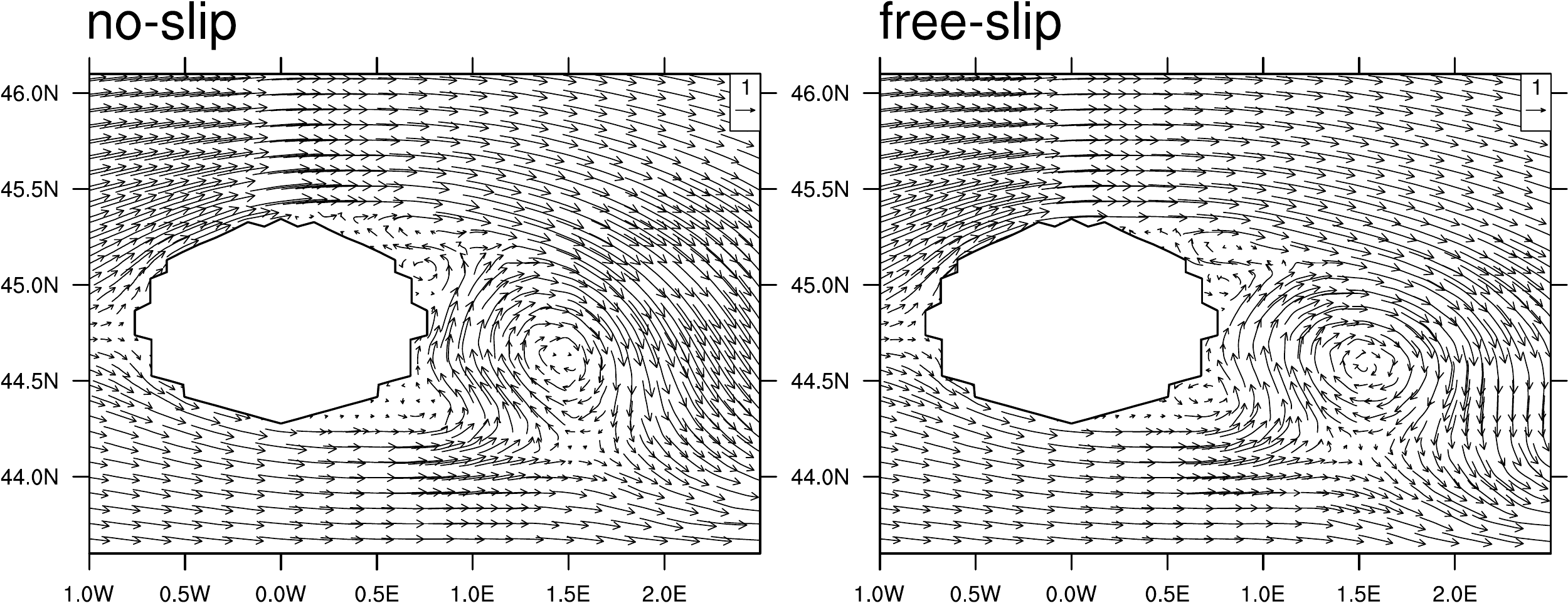} 
 \caption{Vector plot of velocity for the island test with $Re = 3000$ and no-slip (left) or free-slip (right) boundary conditions.}
  \label{bo_fig:dong_vort}
\end{figure}

We obtain different flow regimes when changing viscosity, and therefore the Reynolds number. Analogue to the results in \cite{Dong2007} we obtain two symmetric steady eddies in the lee of the island for $Re \approx 10$, vortices that detach periodically and form a von K\'arm\'an vortex street for $Re \approx 400$ and a fully turbulent behavior in the lee of the island for $Re = 3000$. We will investigate boundary separation in the fully turbulent regime. Figure \ref{bo_fig:dong_vort} shows a snap-shot of the velocity field for no-slip and free-slip boundary conditions.

\section{Criteria for flow separation}
\label{sec:Criteria}

In this section the different criteria for flow separation that are used in this paper are introduced.

\subsection{Prandtl's separation criteria for no-slip boundaries}
\label{sec:Prandtl}

In 1904 Prandtl developed a theory for flow separation of steady flows along no-slip boundaries (\cite{Prandtl1904}). Following Prandtl, separation occurs at a boundary aligned with the $y$ coordinate at a point $P(0,\gamma)$, when the following Eulerian criteria are fulfilled

\begin{align}
\nu \rho \partial_x v(0,\gamma) = 0, \qquad \text{and} \qquad \nu \rho \partial_{xy} v(0,\gamma) < 0, 
\label{bo_prandtl_sp}
\end{align}

where $v(x,y)$ is the meridional velocity, $\nu$ is viscosity and $\rho$ is the density of the fluid. The first condition is a necessary condition that states that the wall shear vanishes, the second condition is a sufficient condition that states that the wall shear admits a negative gradient.

Since $\nu$ and $\rho$ are positive and greater than zero and we want to study general, curved coast lines, we define the two criteria for separation points for general coast lines to be 

\begin{align}
\mathbf{n} \cdot \left[ \left( \nabla \mathbf{u} \right) \cdot \mathbf{t} \right] = 0, \qquad \text{and} \qquad \partial_{\mathbf{t}} \left( \mathbf{n} \cdot \left[ \left( \nabla \mathbf{u} \right) \cdot \mathbf{t} \right] \right) \; < 0, 
\label{bo_prandtl_ge}
\end{align}

where $\mathbf{t}$ is the tangential, and $\mathbf{n}$ is the normal unit vector in respect to the boundary. $\partial_{\mathbf{t}}$ denotes the spatial derivative in the tangential direction, not the time derivative $\partial_{t}$.

\subsection{The topology of boundary separation from no-slip boundaries}
\label{sec:Ghil}

If a boundary current from the south joints a boundary current from the north along a straight western boundary and separates into the domain, the current from the north will make a turn to the left while the current from the south will make a turn to the right. Vorticity will change its sign at the separation point. While a zero point of vorticity might be a necessary condition for separation, it is not a sufficient condition. 

We calculate vorticity as change of the velocity components in tangential and normal direction in respect to the boundary, and evaluate the following Eulerian criterion for possible separation points

\begin{equation}
\label{vorticity}
\omega =  \mathbf{n} \cdot \left[ \left( \nabla \mathbf{u} \right) \cdot \mathbf{t} \right] -  \mathbf{t} \cdot \left[ \left( \nabla \mathbf{u} \right) \cdot \mathbf{n} \right] = 0. 
\end{equation}

Zero vorticity is a very similar criterion to the necessary criterion by Prandtl, since the normal velocity through the boundary is zero and the change of the normal velocity should be zero as well, due to mass conservation at the boundary. 

In a series of papers Ghil, Ma, and Wang (\cite{Ma2001, Ghil2001, Ghil2005}) have studied the topology of flow fields mainly with no-slip, but also with free-slip boundary conditions. They present a theory to determine structural bifurcation of two-dimensional incompressible vector fields. 
An additional paper studies bifurcation points in a flow parallel to the boundary from which a separation and a reattachment point emerge (\cite{Ghil2004}). When the vorticity has been positive in the neighborhood, such a bifurcation point occurs at a time $t^*$ at a point $P(0,\gamma,t^*)$ along a meridional boundary at $x=0$, if the following conditions are satisfied
\begin{align}
 \omega(0,\gamma,t^*) = 0 , \qquad \partial_y \omega(0,\gamma,t^*) &= 0, \qquad \partial_{yy} \omega(0,\gamma,t^*) > 0,\notag \\
\text{and} \qquad \partial_t \omega(0,\gamma,t^*) &< 0. \notag
\end{align}
The conditions stand for a declining vorticity along the boundary that has a local minimum at $\omega = 0$. If the vorticity field in the neighborhood is negative, a bifurcation point is indicated by an increasing vorticity field that has a local maximum at $\omega = 0$. 

The study was done for no-slip boundary conditions and incompressible fluids. There is no mathematical support that 
the results should be valid in the used shallow-water system, since the shallow-water setup is not incompressible. However, the divergence of the velocity field $\nabla \cdot \mathbf{u}$ is typically very small in our experiments. Therefore, we test this criterion within our experimental setup.

\subsection{Haller's separation criteria for no-slip boundaries in unsteady flows}
\label{sec:Haller}

\cite{Haller2004} evaluates the change of the wall shear for flow separation from no-slip boundaries, similar to Prandtl's theory. Haller's theory offers separation criteria for unsteady but fixed flow separation in general two-dimensional velocity fields based on dynamical systems theory. The flow trajectory at the separation point is assumed to be a non-hyperbolic saddle point. Haller combines this assumption with the continuity equation at the boundary, to derive a theory that provides necessary and sufficient conditions for flow separation and high-order approximations for unsteady separation profiles in the vicinity of the boundary.

For a boundary aligned with the $y$-coordinate ($x=0$), a necessary condition for a so-called effective separation point $y = \gamma_{\text{eff}}$ at a given time $t_0$ is
\begin{equation}
 \int\limits_{t_0}^t \frac{\partial_x v(0, \gamma_{\text{eff}}, \tau)}{\rho (0, \gamma_{\text{eff}},\tau )} d \tau = 0.
\label{bo_noslip_sep}
\end{equation}

The quantity is integrated backwards in time ($t < t_0$). The effective separation point will converge to the real separation point $\gamma$ for $t \rightarrow - \infty$

\begin{equation}
  \gamma = \lim_{t \to -\infty} \gamma_{\text{eff}}(t,t_0). \notag
\end{equation}

In the derivation of this criterion, Haller had to assume that the density and the time integrated second derivative $\partial_{xy} u$ remain bounded. 
For zero integration time, the necessary criterion by Haller is equivalent to the necessary condition by Prandtl (equation \eqref{bo_prandtl_sp}), since viscosity and density are always positive and greater than zero.

A sufficient condition for separation which is also derived in \cite{Haller2004}, is

\begin{equation}
 \int\limits_{t_0}^{-\infty} \left[ \frac{\partial_{yx} v(0,\gamma, \tau) - \partial_{xx} u(0,\gamma, \tau)}{\rho (0,\gamma,\tau )} - 2 \partial_{yx} u(0,\gamma, \tau) \int\limits_{t_0}^\tau \frac{\partial_{x} v(0,\gamma, s) }  {\rho (0,\gamma,s )} ds \right] d \tau = \infty.
\label{bo_noslip_sep3}
\end{equation}

It can be shown that this criterion reduces to the sufficient criterion by Prandtl when applied to steady flows (\cite{Haller2004}). In the present model, second order derivatives of the velocity in space are difficult quantities, since they need to be reconstructed with neighboring grid cells.
The two combined time integrations make it very complicated to calculate the sufficient condition \eqref{bo_noslip_sep3}. Furthermore, it is difficult to analyze equation \eqref{bo_noslip_sep3}, since infinite values will not be reached, when model runs are evaluated. For these reasons, we solely use the necessary condition, and evaluate points we call `possible separation points', since these points are not verified by a sufficient condition.

To obtain an applicable separation criterion based on the theory by Haller, for flow separation in a shallow-water model with realistic coast lines, we replace the fluid density in equation \eqref{bo_noslip_sep} with the water depth, and modify the criterion to be

\begin{align}
 \lim_{t\rightarrow - \infty} \int\limits_{t_0}^t \frac{ \mathbf{n} \cdot \left[ \left( \nabla \mathbf{u} \left(x,y,\tau \right) \right) \cdot \mathbf{t} \right] }{H}  d \tau = 0.
\label{bo_haller}
\end{align}


\subsection{Separation on free-slip boundaries for unsteady flows}
\label{sec:HallerLekien}

\cite{Lekien2008} studied separation from \textbf{free-slip} boundaries. They state that flow separation takes place at a point $\mathbf{x}$ situated on the boundary, when the following three assumptions are satisfied:

\begin{enumerate}
\setlength{\itemsep}{+8pt}
 \item $\mathbf{x} \left( t \right)$ attracts other trajectories within the boundary.
 \item $\mathbf{x} \left( t \right)$ has an unique manifold that is uniformly bounded away from a portion of boundary, containing $\mathbf{x} \left( t \right)$ in backward time.
 \item Both of the properties above are robust.
\end{enumerate}

Lekien and Haller evaluated previous work of \cite{Fenichel1971} and \cite{Mane1978} and concluded that the three assumptions above are sufficient and necessary conditions for normal hyperbolicity at $\mathbf{x}  \left( t \right)$, where the flow along the coast line represents the stable and the separated flow into the domain represents the unstable trajectory of $\mathbf{x} \left( t \right)$. Starting here, Lekien and Haller studied the scaling of the velocity field along the coast line, towards the separation point and derived the following two Lyapunov type numbers
\begin{align}
 \lambda_{\mathbf{t}} = & \limsup_{T\rightarrow + \infty} \frac{1}{T} \int \limits_{t-T}^t \left[ \mathbf{t} \cdot \left( \left( \nabla \mathbf{u} \left(x,y,\tau \right) \right) \cdot \mathbf{t} \right) \right]_{\mathbf{x}} d\tau,
 \label{bo_lekien1}\\
\lambda_{\mathbf{n}} = & \liminf_{T\rightarrow + \infty} \frac{1}{T} \int \limits_{t-T}^t \left[  \mathbf{n} \cdot \left( \left( \nabla \mathbf{u} \left(x,y,\tau \right) \right) \cdot \mathbf{n} \right) \right]_{\mathbf{x}} d\tau, 
\label{bo_lekien2}
 \end{align}

where the lowered ${\mathbf{x}}$ shall indicate, that we integrate along a flow trajectory ${\mathbf{x} \left( t \right) }$, and follow an imaginary particle in the flow field along the coast for these Lagrangian criteria. Flow separation takes place whenever $\lambda_{\mathbf{t}} < 0$, and $\lambda_{\mathbf{n}} > 0$. Similar criteria can be proved for flow reattachment (see \cite{Lekien2008}).

The results are derived for free-slip boundary conditions and moving separation points in a continuous fluid. If $\lambda_{\mathbf{t}}$ and $\lambda_{\mathbf{n}}$ are calculated for discrete model output and the flow trajectories are integrated over finite time intervals, $\lambda_{\mathbf{t}}$ and $\lambda_{\mathbf{n}}$ will not be zero at points where no separation takes place. Therefore, we replace the conditions `smaller or greater than zero' with the conditions `local minima and maxima' of the two Lyapunov type quantities, as it was done in \cite{Lekien2008}.

\subsection{The adverse pressure gradient}
\label{sec:presgrad}

\cite{Haidvogel1992} simulated boundary currents in ocean domains with idealized coast lines and observed that boundary separation coincides with a strong adverse pressure gradient for \textbf{no-slip and free-slip} boundary conditions. A pressure gradient opposes the flow along the wall and forces it to separate. \cite{Haidvogel1992} analyzed the flow field in 60 km distance to a straight, meridional coast. 
Since Haidvogel et al. studied the quasigeostrophic potential vorticity equation they could not evaluate the pressure gradient solely, but investigated a quantity they called the higher-order pressure gradient which includes a part of the Coriolis term. The existence of an adverse pressure gradient at boundary separation points is also known from the literature in Fluid Dynamics (\cite{Monin2007}). 
It was shown in \cite{Ghil2004} that an adverse pressure gradient is also present at the bifurcation point discussed in subsection \ref{sec:Ghil}.

Following the results of \cite{Haidvogel1992} we evaluate the Eulerian quantity

\begin{equation}
 	\lambda = \mathbf{t} \cdot \left( \nabla \cdot h \right) \cdot \text{sign}[ \mathbf{t} \cdot \mathbf{u} ]. 
\label{bo_advpress}
\end{equation}
 
$\lambda$ describes the change of the height field in the direction of the flow along the coast. This is equivalent to an adverse pressure gradient. If $\lambda$ shows a maximum at a value larger than a constant $c$, which is adjusted to the flow field, we take this as a criterion for flow separation points.

\section{Results}

In this section, we provide numerical result for the different separation criteria and test cases. No-slip boundaries are studied in Section \ref{steady_no} and \ref{eddy_no} for the idealized coast line and the eddy-resolving test case. Free-slip boundary conditions are studied in Section \ref{steady_free} and \ref{eddy_free} for both test cases.

\subsection{Steady flows with no-slip boundaries and idealized coastlines}
\label{steady_no}

 \begin{figure}[ht!]
    \center
 \hspace{-6.0cm}  a) \hspace{6.0cm} b) \hspace{8.0cm}\\
    \includegraphics[width=0.4 \textwidth, angle=90]{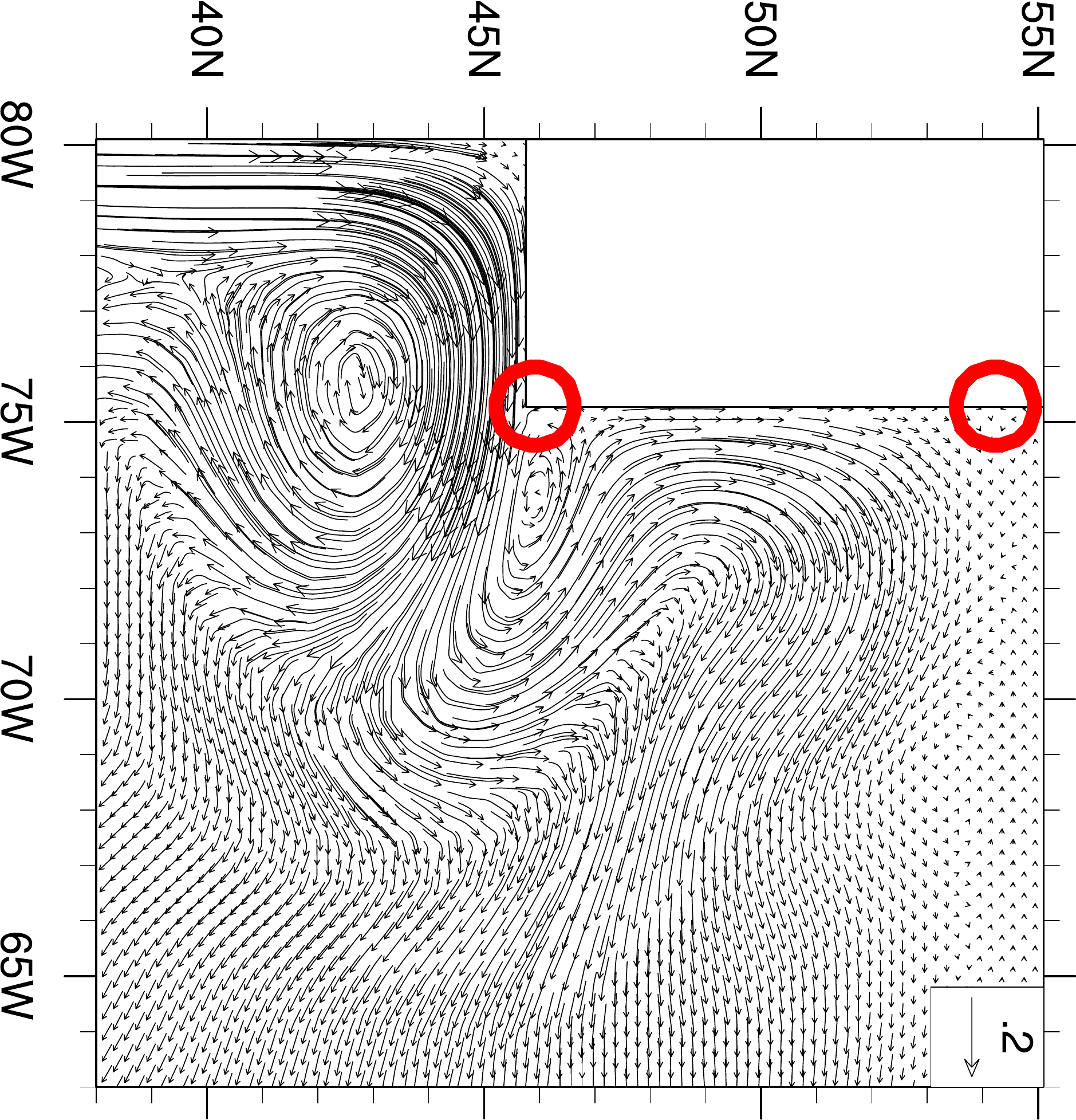} \includegraphics[width=0.4 \textwidth, angle=90]{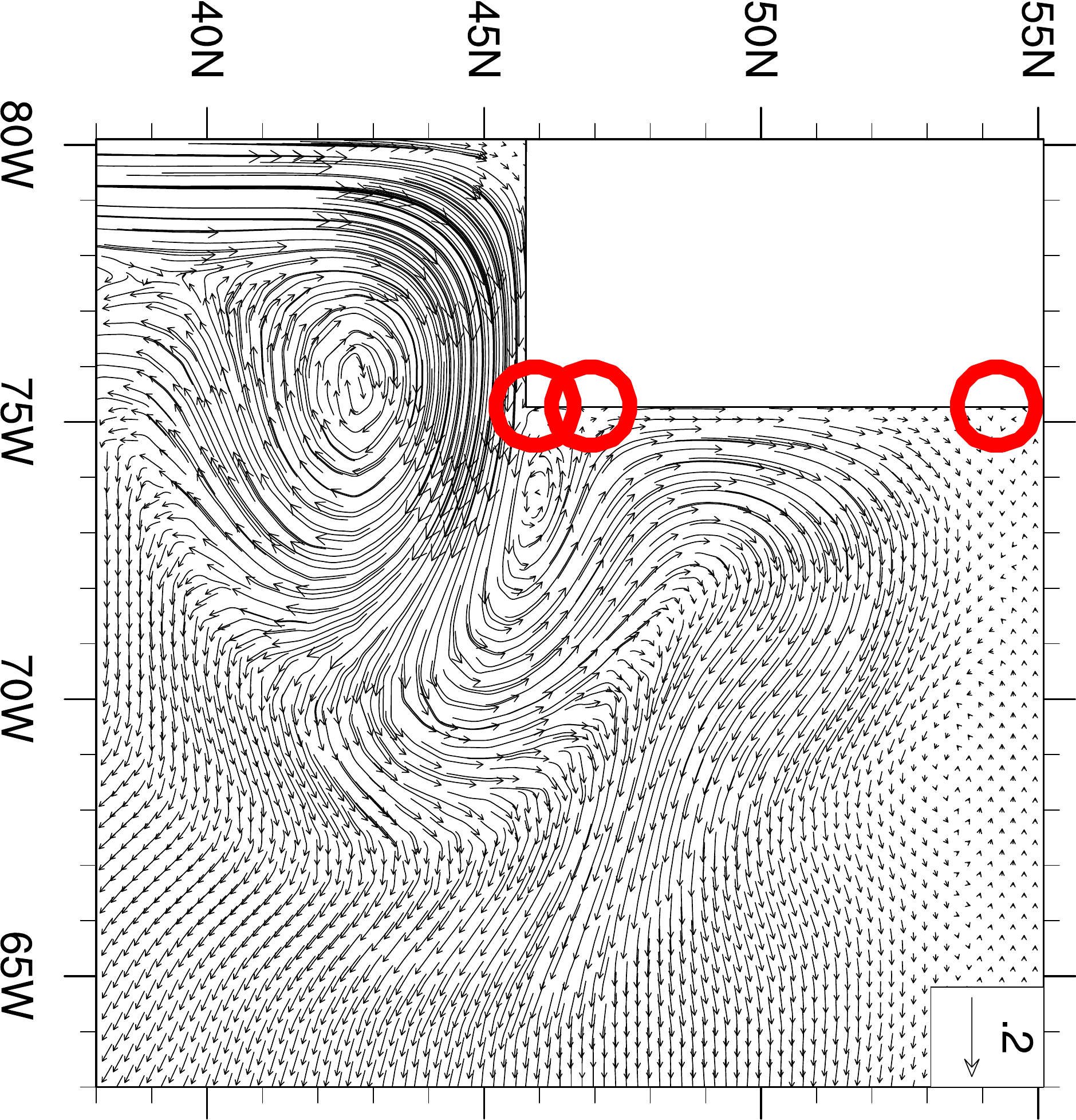} \\
 \hspace{-6.0cm}   c)  \hspace{6.0cm} d) \hspace{8.0cm} \\
    \includegraphics[width=0.4 \textwidth, angle=90]{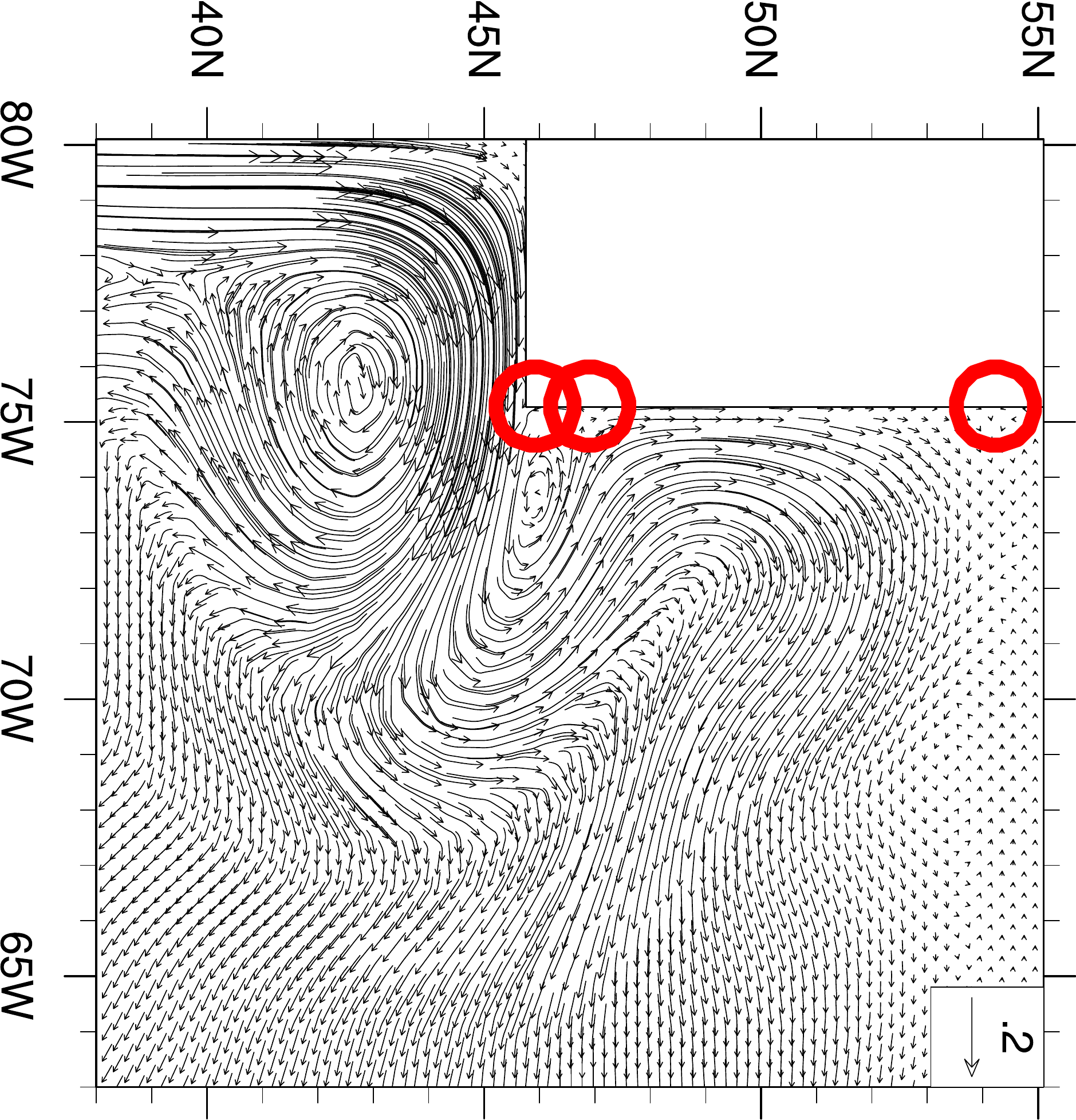} \includegraphics[width=0.4 \textwidth, angle=90]{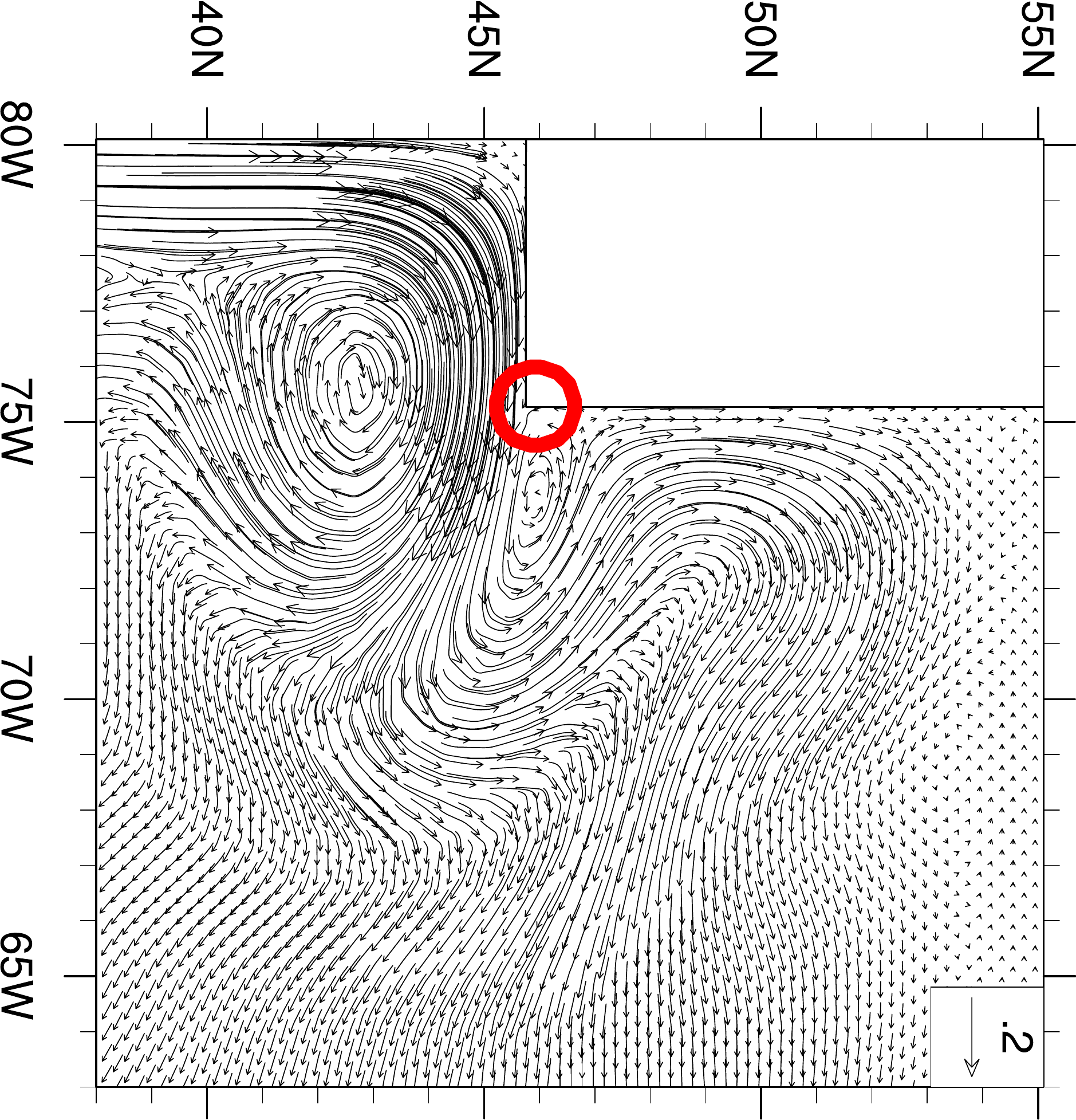} 
     \caption{Separation points detected for the idealized coast line test with no-slip boundaries using a) Prandtl's criteria, b) points of zero vorticity, c) Haller's criterion (integrated for 26 days) and d) the adverse pressure gradient with $c = 5\cdot 10^{-10}$.}
   \label{bo_fig:ideal_noslip}
 \end{figure}
 
Figure \ref{bo_fig:ideal_noslip} shows the results for the detection of separation points in the idealized coastline test on no-slip boundaries. The criteria for separation of Prandtl's theories are able to detect the separation points (compare Figure \ref{bo_fig:ideal_noslip} a with Figure \ref{bo_fig:dengg_fs_jk}).
Points of zero vorticity and Haller's separation criterion provide the same results as Prandtl's criteria. However, the reattachment point is detected as well since the sufficient separation criterion is not evaluated.  
The criterion of an adverse pressure gradient is able to detect the more prominent separation point at the edge of the obstacle, but it fails to detect the second separation point. The necessity to define $c$ for the criterion (see equation \eqref{bo_advpress}) can cause lost detections. If $c$ is too large, not all separation points will be detected. If $c$ is too small, too many points are detected (not shown here). 
In summary, all evaluated separation criteria are able to identify separation points in the idealized coast line test. Separation points appear to be represented reasonably well.

\subsection{Eddy-resolving flows with no-slip boundaries}
\label{eddy_no}

\begin{figure}[ht!]
  \center
        \includegraphics[width=1.0 \textwidth, angle=0]{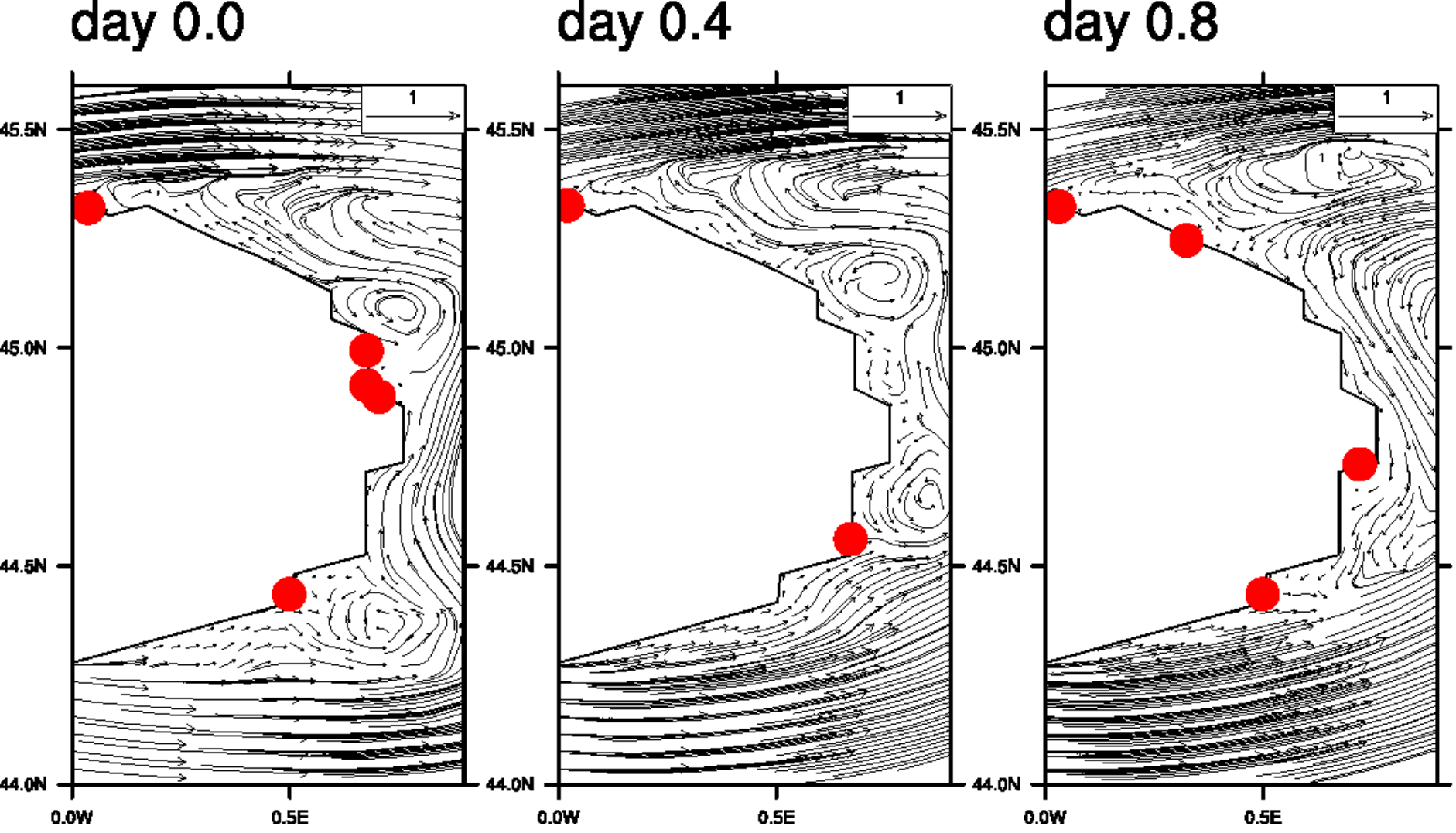} 
    \caption{Separation points detected with Prandtl's separation criteria for the no-slip island test and three time steps.}
   \label{bo_fig:dong_prandtl_ns_600}
 \end{figure}

 \begin{figure}[ht!]
    \center
   \includegraphics[width=1.0 \textwidth, angle=0]{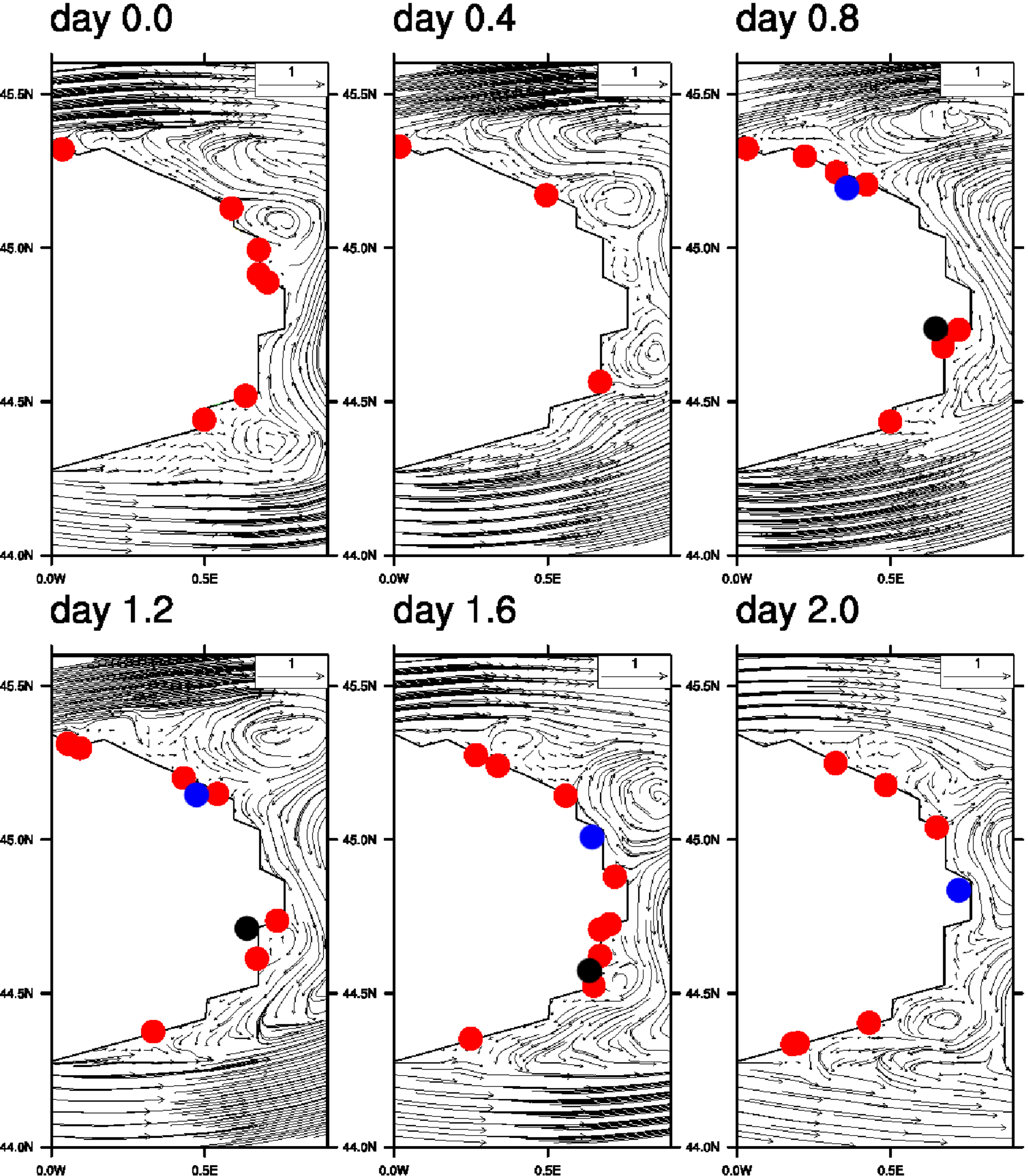} 
     \caption{Points of zero vorticity for the no-slip island test. Red dots mark the detected possible separation points. Dots with other colors indicate the position of specific eddies that move along the coast.}
   \label{bo_fig:dong_ghil_ns}
 \end{figure}

 \begin{figure}[ht!]
    \center
    \includegraphics[width=0.8 \textwidth]{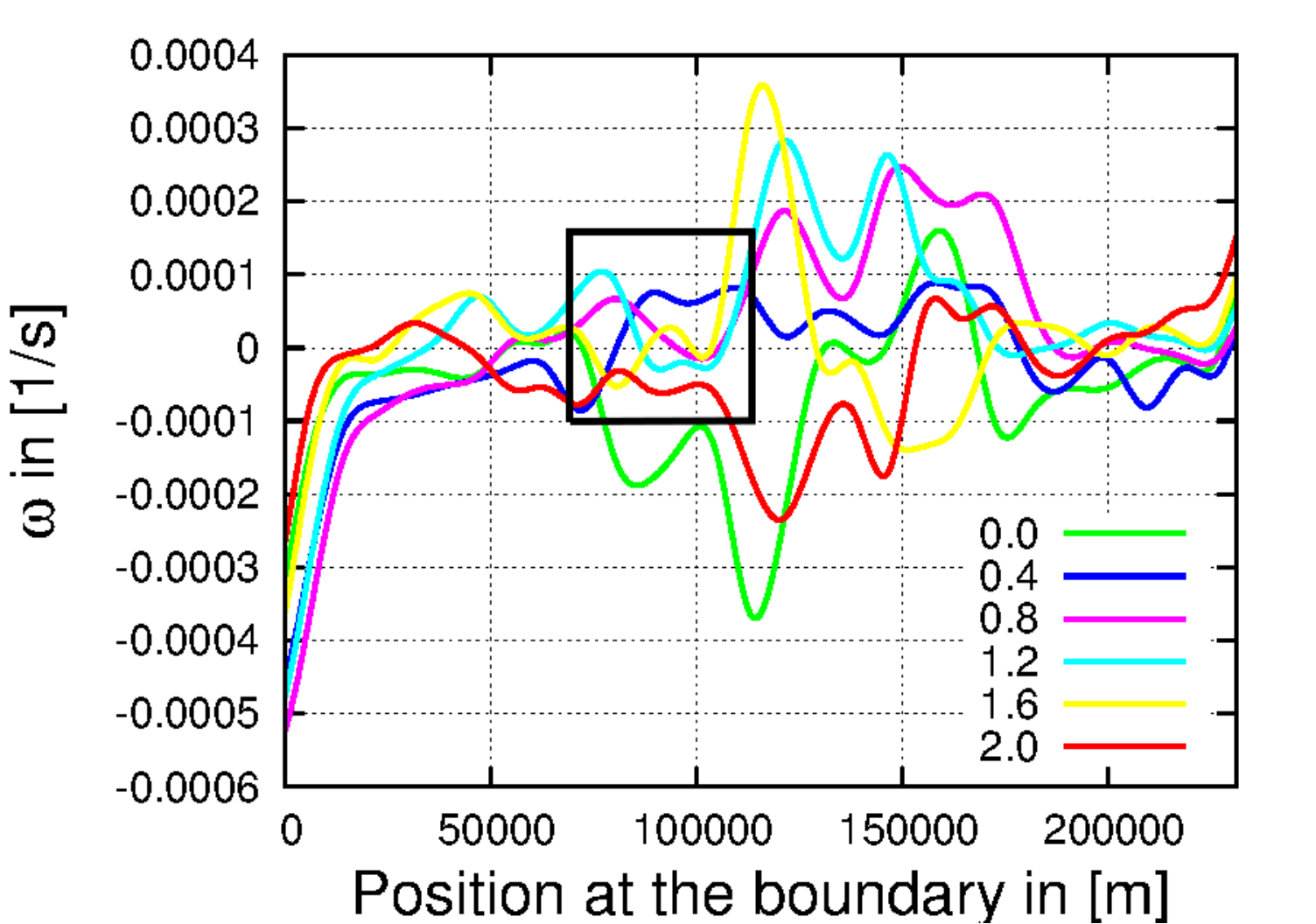} 
     \caption{Vorticity along the boundary for the flow fields plotted in Figure \ref{bo_fig:dong_ghil_ns}, for different time steps (in days). The plotted part of the boundary starts in the south and ends in the north of the visible coast line in Figure \ref{bo_fig:dong_ghil_ns}. The black box marks the area in which the black eddy in Figure \ref{bo_fig:dong_ghil_ns} is born (at day 0.8) and detaches from the coast (at day 2.0).}
   \label{bo_fig:dong_ghil_ns_lambda}
 \end{figure}

Figure \ref{bo_fig:dong_prandtl_ns_600} and \ref{bo_fig:dong_ghil_ns} show the results for the detection of separation points in unsteady flows on no-slip boundaries when applying the criteria for separation of Prandtl's theories and evaluating points of zero vorticity. Both criteria provide reasonable results for the visible separation points in the unsteady test. The zero vorticity criterion can not distinguish between separation and reattachment since no sufficient condition is evaluated. However, the use of the sufficient condition by Prandtl would allow the differentiation between separation and reattachment points.

Figure \ref{bo_fig:dong_ghil_ns_lambda} shows the vorticity field along the coast line of the island for the time steps presented in Figure \ref{bo_fig:dong_ghil_ns}. It is possible to track the position and to identify the generation and separation of eddies along the coast line as predicted in \cite{Ghil2004} by evaluating the change of the position of points of zero vorticity. 
As an example, the birth of a new separation point with the given mechanism by \cite{Ghil2004} can be seen for the eddy indicated by the black dots in Figure \ref{bo_fig:dong_ghil_ns}. The vorticity pattern which is apparent when the eddy is born, is marked by the black box in Figure \ref{bo_fig:dong_ghil_ns_lambda}. 

\begin{figure}[ht!]
  \center
  \includegraphics[width=1.0 \textwidth, angle=0]{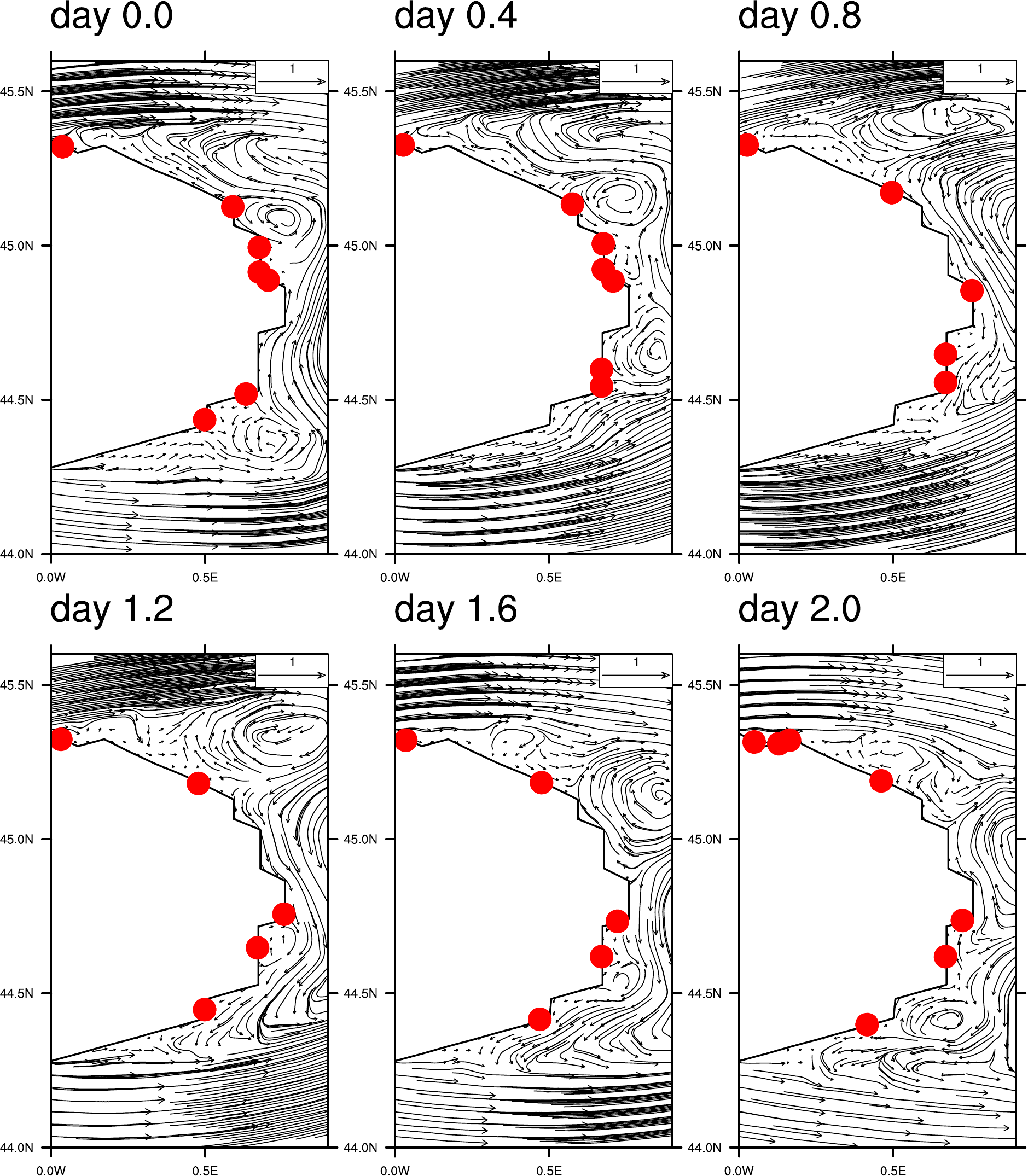} 
      \caption{Separation points detected with the theory by Haller for the no-slip island test. The time integrations for the separation criterion start at day 0 for all plots.}
   \label{bo_fig:dong_haller_ns_600}
 \end{figure}

Figure \ref{bo_fig:dong_haller_ns_600} shows the results for the criterion by Haller for the unsteady island test. It needs to be taken into account that the streamlines that are plotted in the Figure represent Eulerian snapshots of the velocity field and do not show flow trajectories of the unsteady flow field. It is known that separation criteria in the Eulerian framework, such as vanishing wall shear, and material spike formations in the Lagrangian framework can be displaced slightly in unsteady flow fields (see for example \cite{Haller2004} and \cite{Weldon2008}). 
Therefore, the results of Haller's criterion would probably look better if evaluated in the Lagrangian framework and plotted against the actual flow trajectories 
(personal communication George Haller). 
Unfortunately, this kind of information is extremely difficult to obtain within our modeling framework
and we have to make the best out of the Eulerian snapshots.

To investigate the influence of the integration time $t$ on the necessary criterion of Haller's theory, we show several time steps with increasing $t$. At zero integration times, Prandtl's and Haller's criterion are equivalent and numerical results confirm this, with the difference that reattachment points will be detected as well by Haller's criterion since the sufficient criterion is not evaluated (see day 0.0 and compare to Figure \ref{bo_fig:dong_prandtl_ns_600}). However, for all other integration times, results differ significantly between the two criteria while the detected points should only be displaced slightly. The positions of detected separation points for Haller's criterion are not changing much with time, although the topology of the flow field is changing fast (compare day 1.2, day 1.6 and day 2.0). 

 \begin{figure}[ht!]
    \center
     \includegraphics[width=0.3 \textwidth, angle=0]{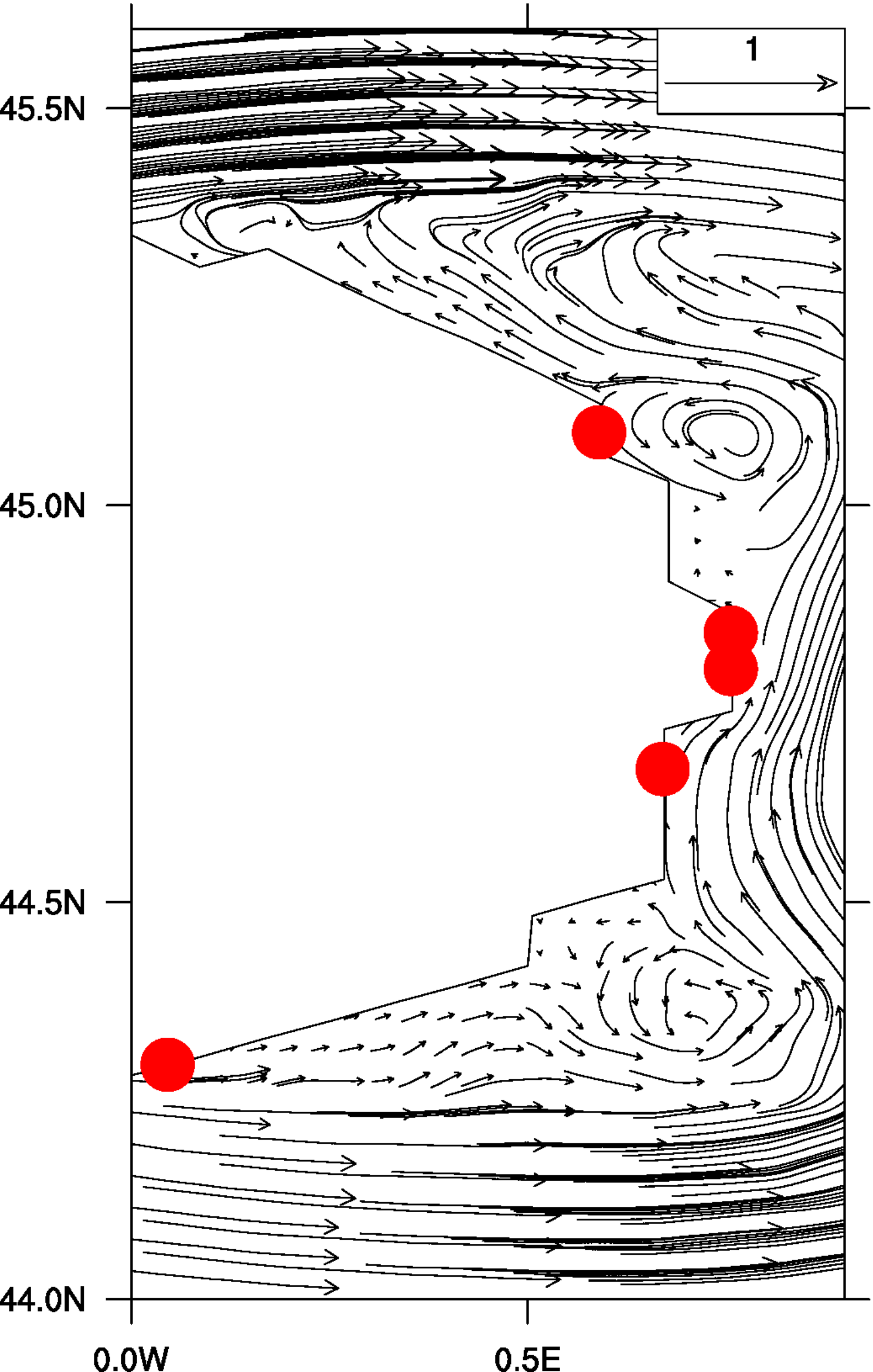}
     \caption{Separation points detected with criterion \eqref{bo_advpress} for the unsteady test case with $c = 5\cdot 10^{-7}$ and no-slip boundary conditions}
   \label{bo_fig:dong_height_ns}
 \end{figure}

Figure \ref{bo_fig:dong_height_ns} shows the results for the adverse pressure criterion. The detected points do not appear unreasonable but many separation points are not detected.
 

\subsection{Steady flows with free-slip boundaries and idealized coastlines}
\label{steady_free}

 \begin{figure}[ht!]
    \center
        \includegraphics[width=0.45 \textwidth, angle=90]{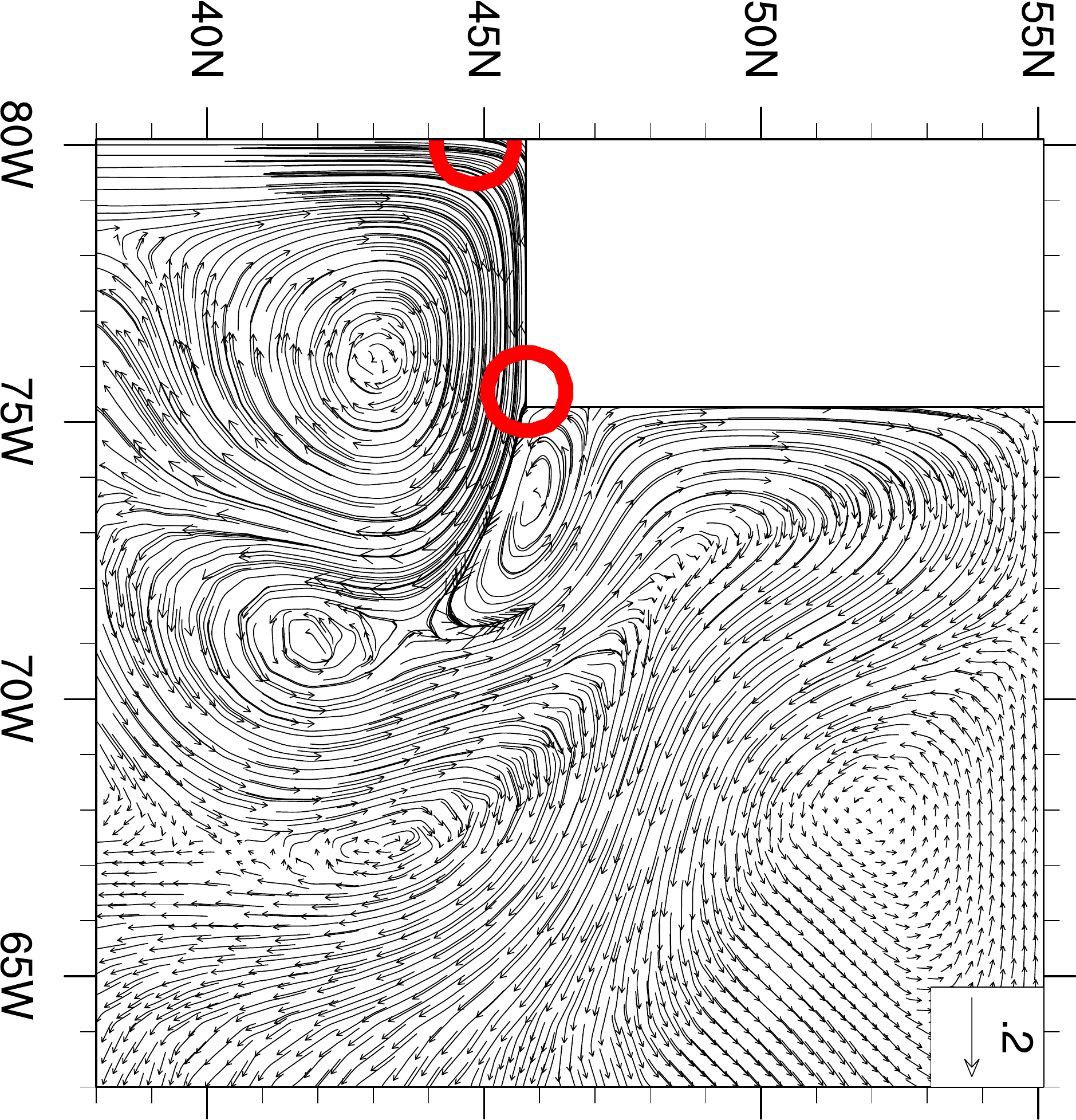}
     \caption{Separation points detected with the criterion of an adverse pressure gradient for the steady test case with $c = 5\cdot 10^{-7}$ and free-slip boundary conditions.}
   \label{bo_fig:dengg_height_ns}
 \end{figure}
 
 \begin{figure}[ht!]
    \center
\includegraphics[width=0.45 \textwidth, angle=90]{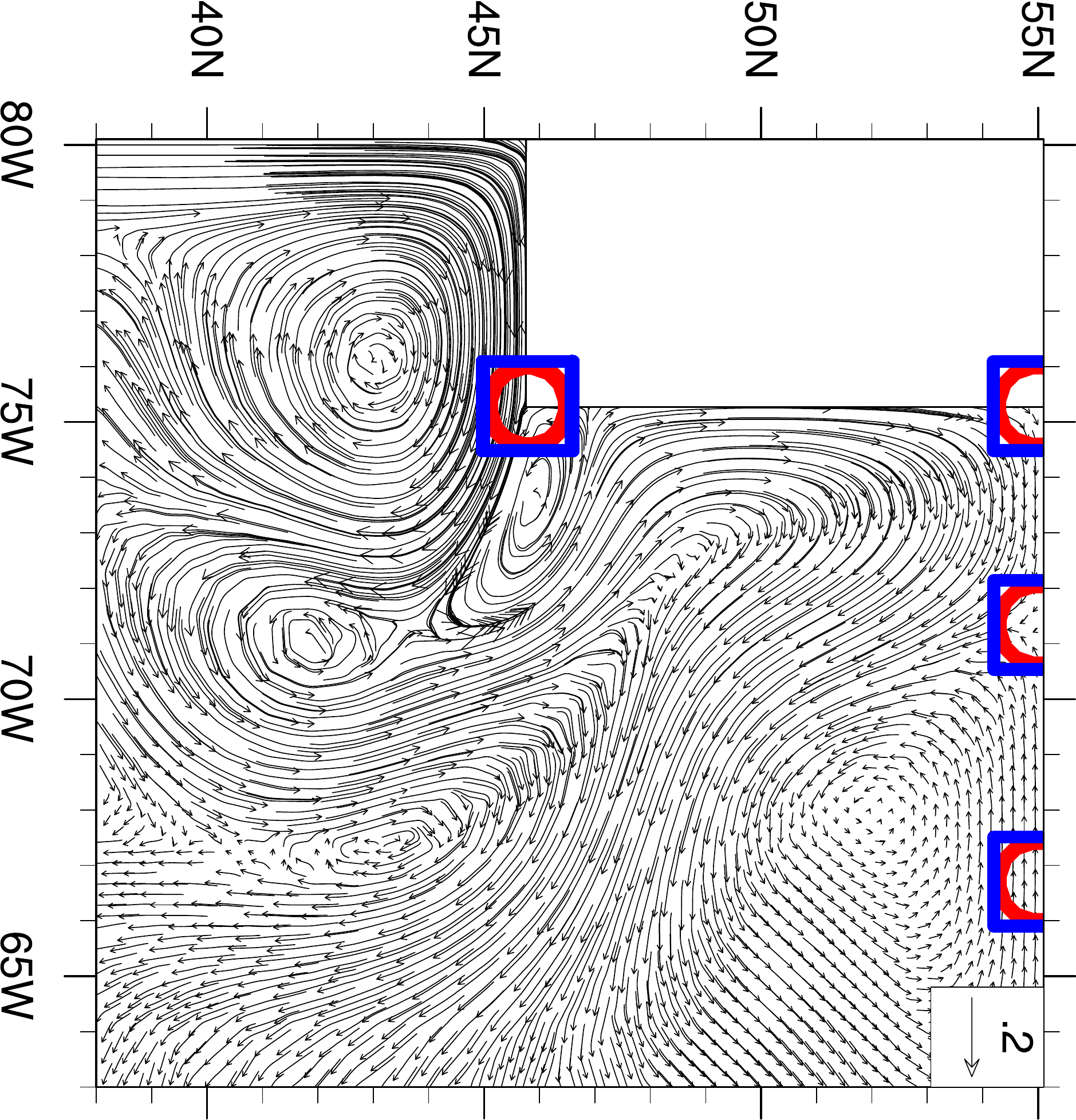}
\includegraphics[width=0.45 \textwidth, angle=90]{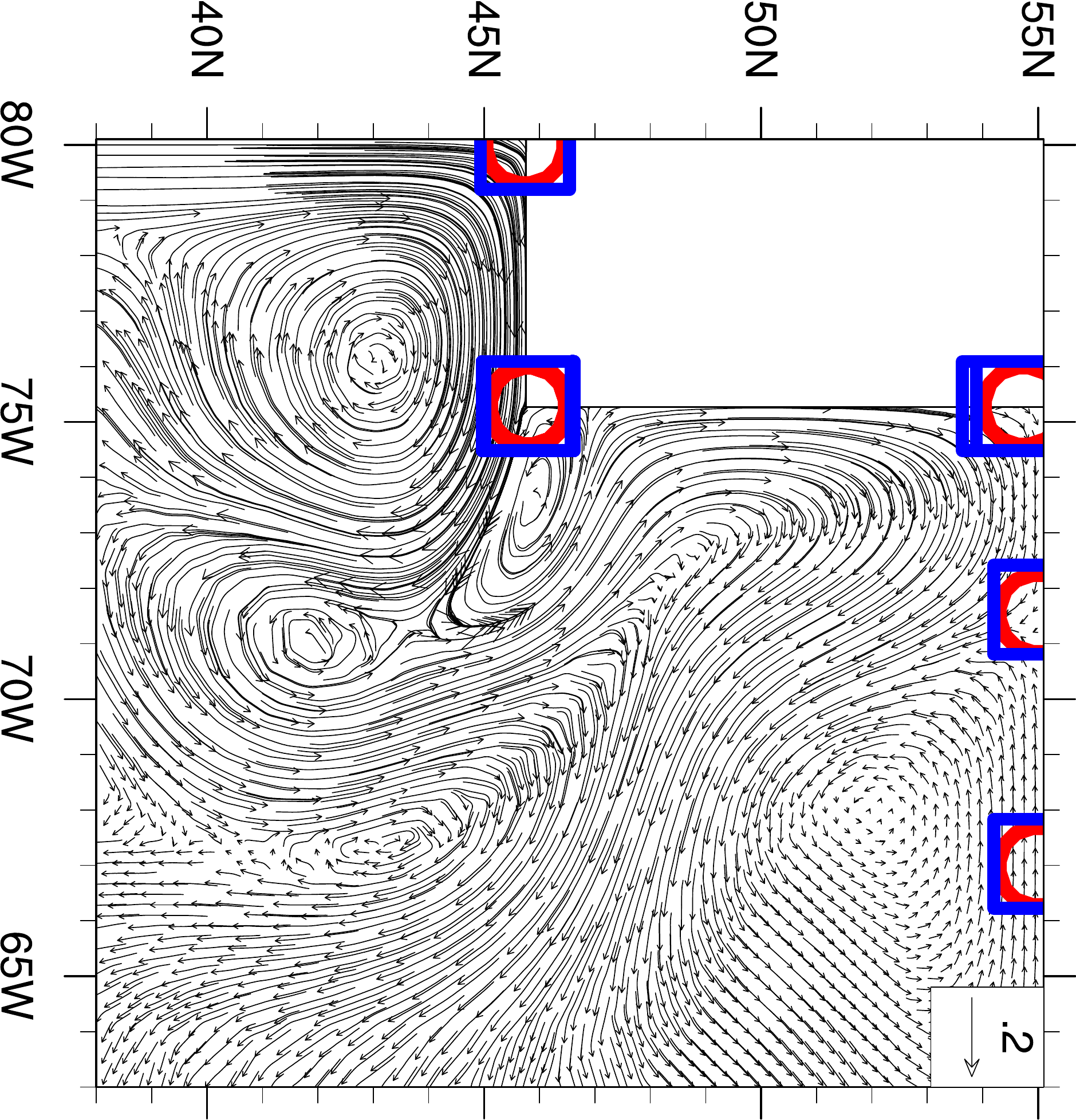} 
        \caption{Separation points detected with the criteria by Lekien and Haller for the free-slip idealized coast line test. The criteria were integrated over 150 (left) and 26 days (right). The red circles mark minima of $\lambda_{\mathbf{t}}$, the blue squares mark maxima of $\lambda_{\mathbf{n}}$. }
   \label{bo_fig:dengg_lekien_fs}
 \end{figure}
 
Figure \ref{bo_fig:dengg_height_ns} and \ref{bo_fig:dengg_lekien_fs} show the results for the idealized coast line test with free-slip boundaries for the adverse pressure gradient criterion and for the criteria by Lekien and Haller. The adverse pressure gradient criterion detects the most prominent separation point correctly but it shows one incorrect detection and fails to detect the second separation point. The criteria of Lekien and Haller are able to identify the separation points (compare with Figure \ref{bo_fig:dengg_fs_jk}), but they appear to be sensitive to turns of the coast line and detect separation points where no separation is visible. Results improve when the integration time is increased.

 \subsection{Eddy-resolving flows with free-slip boundaries}
\label{eddy_free}

 \begin{figure}[ht!]
  \center
\includegraphics[width=1.0 \textwidth, angle=0]{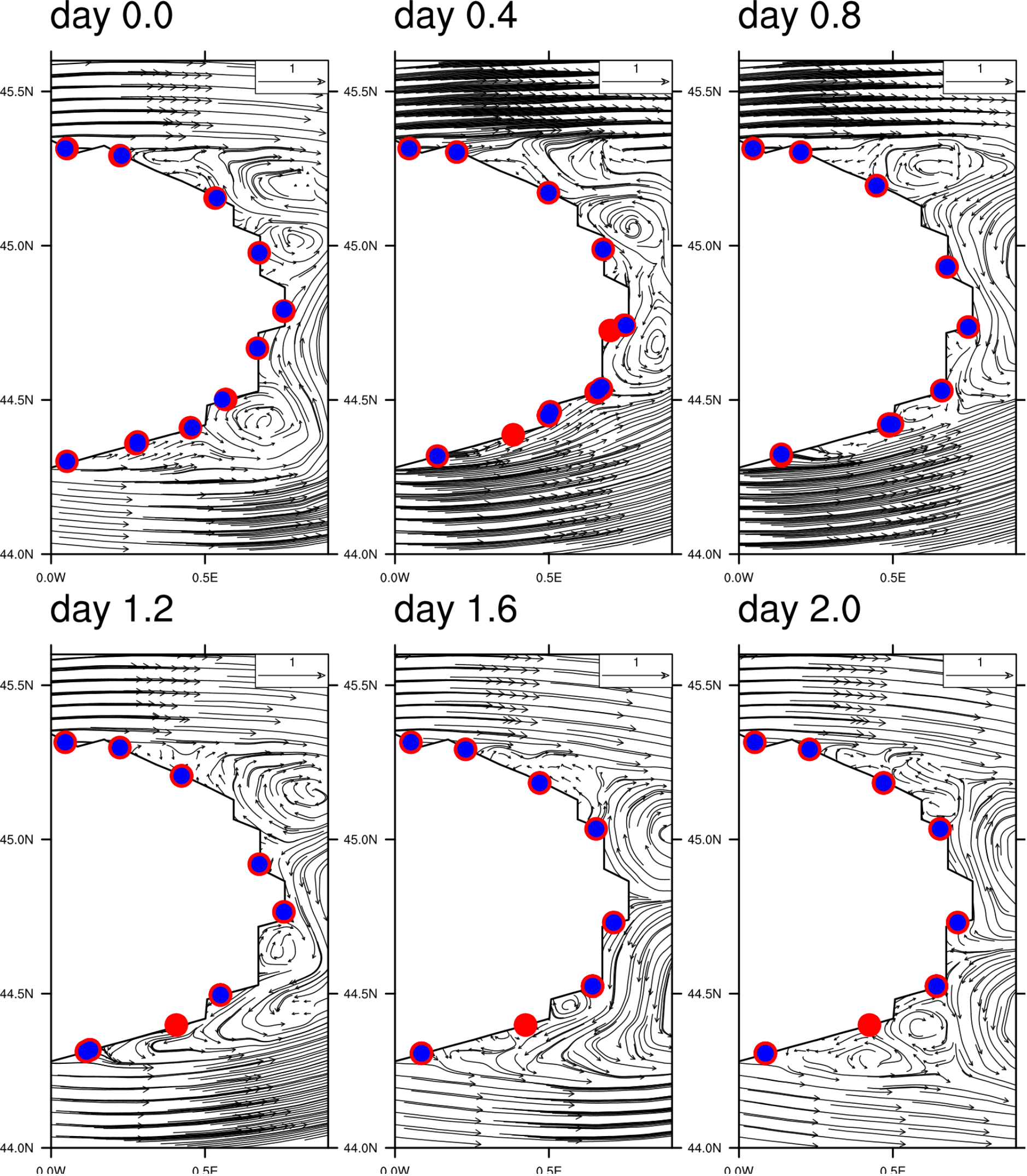} 
    \caption{Separation points detected with the criteria by Lekien and Haller for the free-slip island test. The time integrations for the criteria start at day 0 for all plots. The red dots mark minima of $\lambda_{\mathbf{t}}$, the blue dots mark maxima of $\lambda_{\mathbf{n}}$.}
   \label{bo_fig:dong_lekien_fs_75}
 \end{figure}

The adverse pressure gradient criterion fails to detect separation points for the unsteady flow test case (not shown). Figure \ref{bo_fig:dong_lekien_fs_75} shows the time evolution for the detection of separation points for the criteria of Lekien and Haller. It needs to be mentioned again, that the criteria that are searching for material spikes might look better when plotted against the flow trajectories and not against a velocity snapshot if unsteady flows are considered. 
Several separation points are indicated by the criteria at turns of the coast line for short integration times. 
It is already discussed in \cite{Lekien2008} that it is a weakness of the criteria that a breakaway of the fluid might just be a formation of a local bubble, or a turn of the coast line in our case, and not a large-scale separation. However, since the criteria show a detection of a separation point at almost every grid cell, this does not suggest that separation points are resolved properly. We expected the criteria to improve when integration time is increased. However, for long integration times (day 1.2, day 1.6 and day 2.0), the detected points do not change fast enough with the changing flow field.



\section{Conclusion}

The paper investigates boundary separation points along the coast line of global ocean models. We argue that boundary separation in numerical ocean models differs in essential properties from flow separation in continuous flow fields. In numerical ocean models the coast line will change its direction from one grid cell to its neighbor and the value for viscosity will be selected to be as small as possible at the given resolution; this guarantees that there will be no convergence with resolution and that the coast line and the modeled separation points will be hardly resolved. 
To test if boundary separation points are still represented reasonably well within ocean models along no- or free-slip boundaries despite the coarse resolution, we try to identify boundary separation points using well established criteria for boundary separation.

For no-slip boundary conditions (see section \ref{steady_no} and \ref{eddy_no}) the identification of boundary separation points works reasonably well. All evaluated criteria, namely the criteria by Prandtl, Haller and the adverse pressure gradient, can identify separation points in a test case with idealized coast lines which is not eddy-resolving. For a test with eddy-resolving flow and realistic coast line, the criteria by Prandtl provide reasonable results and it is possible to identify eddies that form and separate along the coastline when evaluating vorticity following the study of flow topology by Ghil et al. (\cite{Ghil2004}). 
This indicates that boundary separation points are represented reasonably well within the model simulations. However, the adverse pressure criterion does not allow a meaningful identification of boundary separation points and the separation criterion based on dynamical systems theory from \cite{Haller2004} provides results that are significantly different to results with Prandtl's separation criteria (except if the evaluation time of Haller's criterion is zero, here results are identical by definition). Haller's criterion fails to adjust quickly enough to the changing flow topology. This indicates that resolution is not sufficient for a correct representation of the full boundary separation profile.

Along free-slip boundaries, both separation criteria, namely the adverse pressure gradient criterion and the criteria from \cite{Lekien2008}, do not provide satisfying results for the eddy-resolving testcase with realistic coastline. Arguably, the identification of boundary separation points is more complicated for free-slip compared to no-slip boundary conditions since the flow is not pushed away from the coast line by the boundary condition. Therefore, separation criteria suffer from detections of local separation points at turns of coast lines that are no real boundary separation points (see Figure \ref{bo_fixpoint_types}). 
But even if turns of the coast line are accepted as boundary separation points, separation criteria suggest separation points at almost every turn of the coast line and therefore at almost every grid cell in an unstructured grid. This is not an indicator for a correct representation of separation. We can assume that the Lagrangian criteria by \cite{Lekien2008} would provide better results if evaluated at higher resolution and the same level of viscosity. Therefore, we conclude that the representation of boundary separation points for free-slip boundary conditions is poor. 

Our results suggest that the representation of flow separation in ocean models is more realistic for no-slip compared to free-slip boundaries at the give level of resolution and viscosity. However, this does not necessarily indicate that no-slip boundary conditions are the better choice for global eddy-resolving ocean models. 
The results of the paper certainly indicate that the limited quality of the representation of boundary separation points due to limited resolution will have an impact on boundary separation. Especially if a boundary current is actually following the coastline within one layer of gridcells. This should be considered in the discussion of the separation of boundary currents in global ocean models. The representation of coast lines and boundary conditions, and therefore the representation of separation points, will be even worse in most existing ocean models compared to results shown in this paper, especially in finite difference models.

At the beginning of this study, our main objective was to identify separation points of boundary currents, such as the Gulf stream, within a running ocean model. This could have been useful for parametrization and adaptive mesh refinement. However, we found that it is very hard to distinguish between a local flow separation, for example at a turn of a coast line or at a local eddy, and a separation of a boundary current with global scale in an eddy resolving model. A distinction between local and global separation will become more and more problematic as resolution increases in ocean models, since the number of small scale eddies along the coastline increases.

\subsection*{Acknowledgments}
We thank David Marshall for very useful comments on earlier versions of this paper, and Stephen Wiggins for a very useful meeting at the beginning of this study.



\bibliographystyle{alpha} 
\bibliography{./paper.bib}





\end{document}